\documentclass[1p,10pt,times]{elsarticle}
\usepackage{amsmath}
\usepackage{hyperref}
\usepackage[switch]{lineno}
\graphicspath{{./images/}}
\usepackage{my}

\begin{document}
\baselineskip11pt

\begin{frontmatter}

\title{Masked BRep Autoencoder via Hierarchical Graph Transformer}

\author[one]{Yifei Li} %\ead{lyf135266@mail.ustc.edu.cn}
\author[one]{Kang Wu} %\ead{kang910042009@gmail.com} 
\author[zero]{Wenming Wu} 
\author[one]{Xiao-Ming Fu} %\ead{fuxm@ustc.edu.cn}
%\cortext[cor]{Corresponding author}

\address[one]{University of Science and Technology of China}
\address[zero]{Hefei University of Technology}

% \author{First Author}
% \cortext[mycorresponingauthor]{Corresponding author}
% \ead{first@firstuni.edu}
% \author{Second Author}
% \address{Department of Mathematics, Stanford University}
% \author{Third Author}
% \address{Department of Computer Science, Tel Aviv University}

% \author{Important: the names of the authors, as well as other identity-revealing information (e.g., funding bodies), should be removed from the original submission to comply with the double-blind review process}

\begin{abstract}
  We introduce a novel self-supervised learning framework that automatically learns representations from input computer-aided design (CAD) models for downstream tasks, including part classification, modeling segmentation, and machining feature recognition.
To train our network, we construct a large-scale, unlabeled dataset of boundary representation (BRep) models.
The success of our algorithm relies on two key components. 
The first is a masked graph autoencoder that reconstructs randomly masked geometries and attributes of BReps for representation learning to enhance the generalization.
The second is a hierarchical graph Transformer architecture that elegantly fuses global and local learning by a cross-scale mutual attention block to model long-range geometric dependencies and a graph neural network block to aggregate local topological information.
After training the autoencoder, we replace its decoder with a task-specific network trained on a small amount of labeled data for downstream tasks. 
We conduct experiments on various tasks and achieve high performance, even with a small amount of labeled data, demonstrating the practicality and generalizability of our model.
Compared to other methods, our model performs significantly better on downstream tasks with the same amount of training data, particularly when the training data is very limited.
\end{abstract}
  
\end{frontmatter}

\section{Introduction}
\label{sec:intro}
As a core component of modern digital industrial manufacturing, computer-aided design (CAD) enhances process efficiency by using computers to create, modify, analyze, and optimize product models, thereby laying a crucial foundation across industries.
In addition to creating models, CAD involves a range of critical tasks, such as machining feature recognition, face segmentation, and shape classification~\cite{shi2020manufacturing,zhang2022machining,yao2023machining,wu2024aagnet,jones2023self,lou2023brep,hou2023fus,zou2025bringing,dai2025brepformer}.

Machine learning for processing CAD models has emerged as a leading research focus, driven by their superior generalizability and efficiency compared to traditional handcrafted methods~\cite{jayaraman2021uv,lambourne2021brepnet}.
In particular, since boundary representations (BReps) are the dominant representation for CAD models due to their precise geometric description, we focus on learning algorithms for BReps.

However, the practical value of learning-based methods is limited by the inherent characteristics of CAD data.
A key characteristic is the scarcity of models, as they often contain sensitive commercial intellectual property that enterprises closely guard.
Hence, constructing large-scale, high-quality labeled datasets is challenging, consequently limiting the performance of learning algorithms.

To address this limitation, a viable way is to perform representation learning on large-scale unlabeled datasets, enabling artificial intelligence (AI) models to automatically extract meaningful features for downstream tasks, thereby improving their performance with limited labeled data.
However, it faces two main challenges.
First, the combination of discrete topology with continuous geometry in BReps complicates the effective handling. % of such heterogeneous data.
Second, the inconsistent and hierarchical nature of BRep graph structures hinders understanding and analysis. %makes it difficult to understand and analyze such complex data.

% %以前的方法的问题
% Many learning methods have been proposed for processing CAD models~\cite{zhang2024brepmfr, wu2024aagnet, zheng2025sfrgnn, zou2025bringing, colligan2022hierarchical, jayaraman2021uv, jones2023self, lou2023brep}. 
% %
% Some of them require a large amount of annotated data for training~\cite{colligan2022hierarchical, zou2025bringing, zhang2024brepmfr, wu2024aagnet, zheng2025sfrgnn}, and most are limited to solving specific downstream tasks, such as feature recognition and surface segmentation.
% %
% Thus, the practicality is limited.
% %
% Other works~\cite{jayaraman2021uv, jones2023self, lou2023brep} are trained using unlabeled data, but the used graph neural network (GNN) encoder is only capable of aggregating local adjacent information while ignoring global information, making it difficult to learn an effective representation from BReps with a complex data structure.

Many learning methods have been proposed for processing CAD models~\cite{zhang2024brepmfr, wu2024aagnet, zheng2025sfrgnn, zou2025bringing, colligan2022hierarchical, jayaraman2021uv, jones2023self, lou2023brep}. 
Some require a large amount of annotated data~\cite{colligan2022hierarchical, zou2025bringing, zhang2024brepmfr, wu2024aagnet, zheng2025sfrgnn} and are limited to specific downstream tasks, such as feature recognition and surface segmentation. 
Other works~\cite{jayaraman2021uv, jones2023self, lou2023brep, Yao2026BRepMAE} train on unlabeled data, but their graph neural network (GNN) encoders only aggregate local information, ignoring global dependencies. %\todo{cite Yao}
%这里是否增加更多Transformer-based architectures的引用？
Recently, Transformer-based architectures have emerged to address this~\cite{zhang2024brepmfr,zou2025bringing}. Especially, BRepFormer~\cite{dai2025brepformer} successfully applies generic graph Transformers to capture long-range global dependencies.
%extreme scale variations具体指的是什么？ 不同面尺度不一样，大的平面与小的倒角面如果共用一套CNN参数效果很差  is easily overwhelmed by the severe data redundancy？这些表述是否严谨？
However, their flat, single-level attention block is ineffective for extreme scale variations (e.g., large planes vs. small fillets) in CAD models and is easily overwhelmed by the severe data redundancy (e.g., planar regions require fewer samples) introduced by dense geometric sampling, leading to inefficient representation learning.

% %我们的方法
In this paper, we introduce a self-supervised learning framework for acquiring effective representations of BReps from a large-scale, unlabeled dataset.
There are two essential technical components.
First, the framework inherits from the elegant masked autoencoder~\cite{he2022masked}.
By representing the BRep as a geometric attributed adjacency graph (gAAG)~\cite{wu2024aagnet}, it learns to reconstruct randomly masked face/edge geometries and their attributes from input CAD models, while keeping the graph topology invariant to enable effective representation learning.
% %
Second, we propose a tailored hierarchical graph Transformer, serving as its core encoder, with both global and local learning to tackle the heterogeneity and extreme scale variations of practical BRep data.
Its architecture centers on a progressive, coarse-to-fine cross-attention model that processes CAD face geometries via dual-resolution UV grids to capture long-range geometric dependencies, followed by a message-passing neural network (MPNN) to aggregate local topological information.
In summary, the progressive abstraction is coupled with the self-supervised masked reconstruction task that guides the encoder to distill essential CAD topologies and geometries rather than merely memorizing raw sampling coordinates.

Extensive experiments on a dataset of 283,018 models demonstrate that our pre-trained hierarchical encoder achieves superior generalization. It significantly outperforms existing representative methods across multiple downstream tasks, especially under highly limited supervision (e.g., $0.1\%$ labeled data).

\section{Related Work}\label{sec:related-work}

\paragraph{Learning for BRep}
The diverse geometries and complex topologies of BReps pose significant challenges for neural network learning.
Methods~\cite{zhang2024point,lei2022mfpointnet,yao2023machining,zhang2022machining,guo2022implicit,zhang2018featurenet} circumvent these challenges by converting BRep models into alternative 3D representations, like point clouds, voxels, or meshes, but fail to exploit the topological information inherent in BReps. %, thereby losing critical information.
A more faithful way treats BRep as a structured graph~\cite{jones2021automate,lambourne2021brepnet,bian2022material,willis2022joinable}, where each BRep face is represented as a node and each BRep edge as a graph edge. 
%propose a method  that
UVNet~\cite{jayaraman2021uv} samples geometric signals from both surfaces and curves %, such as UV grids, which are processed using convolutional neural networks inspired by image understanding, 
to convert BRep models into graphs and applies GNN to aggregate information across the BRep topology.
%
%Then, the GNN is applied to aggregate information across the BRep topology.
%A GNN is then applied to propagate information across the BRep topology by treating edges as message-passing links between face nodes. 
%
Owing to its effectiveness in handling BRep geometry and topology, UVNet has inspired a series of follow-up works~\cite{zhang2024novel,shi2020manufacturing,ning2023part,liu2022supervised,wu2024aagnet,zhang2024brepmfr,wang2023hybrid1,zheng2025sfrgnn,li2025classification}. 
Among them, Wu et al.~\cite{wu2024aagnet} incorporate attributes to achieve improved performance. %, particularly in the machining feature recognition task. enhances the original architecture by 
Zou et al.~\cite{zou2025bringing} leverage Transformer-based attention to capture long-range dependencies across the topology, but they do not use the graph-structured messaging mechanism to aggregate neighborhood information.
%
% In contrast, our method not only utilizes the Transformer to capture global information but also leverages the graph structure to transfer local information, enabling our network to learn an effective representation for BRep.
Moreover, BRepMFR~\cite{zhang2024brepmfr} and BRepFormer~\cite{dai2025brepformer} introduce graph Transformers for global interactions. 
However, applying single-level attention uniformly remains inherently challenging for CAD models' extreme scale variations, as it struggles to decouple macroscopic topologies from dense microscopic details. 
%. Our Cross-Scale Mutual Attention (CSMA) block establishes a dual-resolution ($3\times3$ and $13\times13$) information bottleneck to explicitly decouple macro-anchors from micro-features. 
To address this, we propose a hierarchical architecture featuring a cross-scale mutual attention block that establishes a dual-resolution information bottleneck, explicitly decoupling macro-anchors from micro-features.
Finally, an MPNN aggregates local graph topology, seamlessly synergizing global hierarchical abstraction with local messaging for robust representation learning.

\paragraph{Self-supervised learning for CAD models}
%Self-supervised learning has become a key paradigm in modern deep learning, enabling models to learn meaningful representations without relying on labeled data. 
%
%This approach is particularly appealing for CAD models, where publicly available datasets are limited and high-quality, annotated examples are even scarcer.
Self-supervised learning is especially valuable for CAD models, where publicly available, well-annotated datasets are scarce.
Jones et al.~\cite{jones2023self} propose a reconstruction-based self-supervised learning method that, while effective in few-shot settings, is limited to CAD models composed of basic geometric primitives.
BRep-BERT~\cite{lou2023brep} converts BRep entities into discrete tokens via a GNN tokenizer and trains a Transformer with a masked entity modeling task. However, this discretization step introduces a significant bottleneck by compressing geometries and topologies into a limited number of token IDs.
BRepGen~\cite{xu2024brepgen} uses a hierarchical tree representation to encode BRep models into a latent space, which cannot be effectively used for downstream tasks.
Zhang et al.~\cite{zhang2024point} operate on point clouds, which inherently lack topological and detailed geometric information. %, thus limiting the network's generality.
%take point clouds as the inputs, missing topological and detailed geometric information, thereby limiting the generality of the network.
%
%\todo{discuss Yao}
%
Our method builds upon the masked autoencoder (MAE)~\cite{he2022masked} framework. % to enable effective self-supervised training of a hierarchical graph Transformer (GT) encoder. 
Closely related to our work, Yao et al.~\cite{Yao2026BRepMAE} successfully apply MAE to BRep models by masking latent features and employing an MPNN.
In contrast, we enforce a stricter information bottleneck by directly masking raw geometric inputs and use a hierarchical GT to capture long-range global dependencies prior to local topology aggregation.

\paragraph{Graph Transformer}
%GTs have recently emerged as powerful architectures for modeling graph-structured data. 
%
Unlike traditional GNNs, which rely solely on local message passing, GTs also use attention mechanisms to capture long-range dependencies, thereby becoming powerful architectures for graph-structured data to achieve more expressive and flexible representations.
%Hence, they .
%
Most prior work is centered on molecular property prediction, where input graphs represent chemical compounds, such as Graph-BERT~\cite{zhang2020graphbert}, GROVER~\cite{rong2020self}, GraphiT~\cite{mialon2021graphit}, GraphTrans~\cite{wu2021graphtrans}, SAN~\cite{kreuzer2021sab}, Graphormer~\cite{ying2021graphormer}, and Sgformer~\cite{wu2023sgformer}. 
%
%For instance, Graph-BERT~\cite{zhang2020graphbert} incorporates structural priors such as Weisfeiler–Lehman codes and hop distances to capture topologies without relying on explicit edge features. 
%
%GROVER~\cite{rong2020self} further integrates GNNs and Transformers within a multi-task self-supervised learning framework. 
%
%Other variants, such as GraphiT~\cite{mialon2021graphit}, GraphTrans~\cite{wu2021graphtrans}, SAN~\cite{kreuzer2021sab}, Graphormer~\cite{ying2021graphormer}, and Sgformer~\cite{wu2023sgformer}, improve structural encoding through enhanced positional embeddings or hybrid attention mechanisms.
%
% \tr{Nevertheless, applying GTs to BRep learning remains largely unexplored.
% %
% Since BReps naturally form graphs, GTs are well-suited for learning their representations. % with geometries and topologies
% %
% Hence, we employ a GT encoder tailored to BReps, enabling effective and transferable representation learning across diverse downstream tasks.}
While GTs have shown immense potential, their direct application to BRep data requires careful architectural design. 
Although the pioneers~\cite{zhang2024brepmfr, dai2025brepformer} have introduced GTs to BReps, they predominantly inherit flat attention structures from general-purpose GTs, which limits their effectiveness on complex CAD topologies.
To bridge this gap, we propose a hierarchical GT encoder tailored to BReps. 
By seamlessly integrating a dual-resolution information bottleneck with local message passing, our architecture overcomes the limitations of flat attention, enabling highly expressive and transferable representation learning across diverse downstream tasks.

\section{Method}
\begin{figure*}[t]
  \centering
  \includegraphics[width=0.99\textwidth]{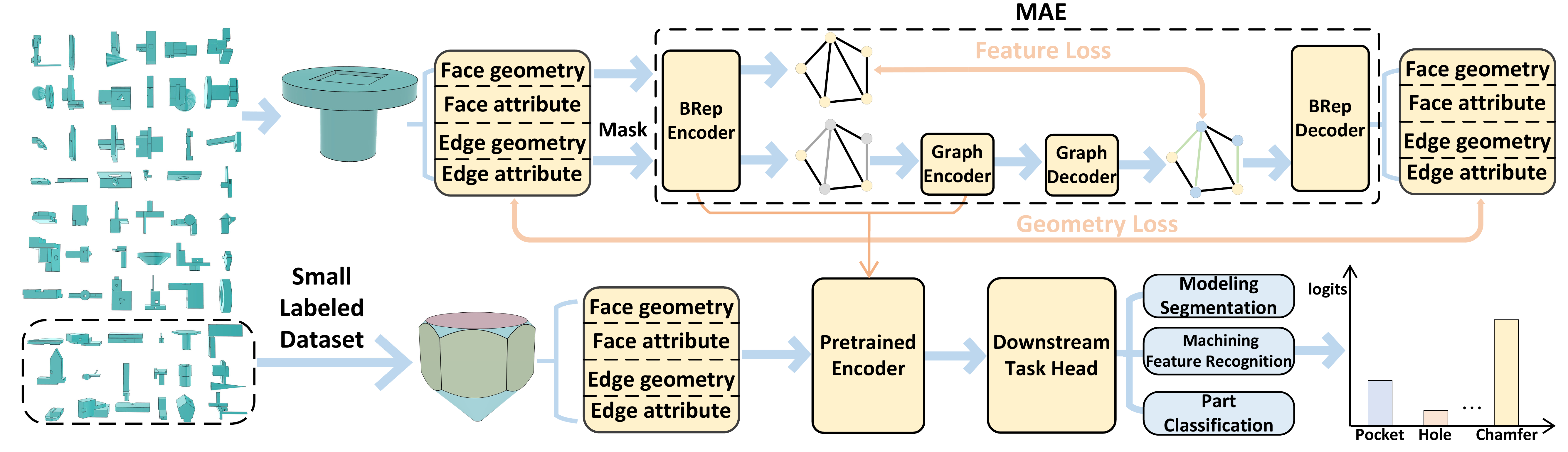}
  \vspace{-2mm}
  \caption{
  %The pipeline of our method. 
  %
  Given a CAD model, our method first applies a BRep encoder to construct a gAAG from the extracted BRep information.
  A Graph encoder is then implemented to update the graph features with both global and local information.
  During pre-training, we train an MAE by reconstructing the randomly masked faces and edges for BRep representation learning (the upper part of the figure).
  In the fine-tuning stage, we train a new network, formed by connecting a task-specific head behind the encoder, using a small amount of labeled data with different loss functions (lower part of the figure).
  %
  %The newly formed network is then trained on a small amount of labeled data with different loss functions (lower part of the figure).
  }
  \label{fig:pipeline}
  \vspace{-1mm}
\end{figure*}

%\paragraph{{Overview}
Our proposed self-supervised BRep representation learning method consists of two parts: BRep information extraction (Sec.~\ref{sec:constructing-gAAG}) and pre-training (Sec.~\ref{sec:pre-training}).
%
%In BRep information extraction, our method uniformly samples on BRep faces and edges to extract the geometric and attribute information (Sec.~\ref{sec:constructing-gAAG}).
%
%Then, in pre-training, we randomly mask the geometric information of BRep models and use our autoencoder to reconstruct them to learn a representation (Sec.~\ref{sec:pre-training}).
For pre-training, we employ the MAE framework to reconstruct the masked geometries in the input models and propose a tailored hierarchical graph Transformer encoder with a cross-scale mutual attention block.
After pre-training, we concatenate a specific-task network behind our encoder to form a new network, and then train it with a small amount of labeled data for various downstream tasks (Sec.~\ref{sec:domain-adaptive}).
The pipeline of our method is shown in Fig.~\ref{fig:pipeline}.

\subsection{BRep information extraction}
\label{sec:constructing-gAAG}
%\tr{To handle the extreme scale variations of CAD  (e.g., ) and alleviate data redundancy on surfaces (e.g., ), we propose a dual-resolution geometric extraction strategy that explicitly decouples global surface trends from local high-frequency details.}
To address extreme scale variations (e.g., large planes vs. small fillets) and reduce surface data redundancy (e.g., planar regions that require fewer sample points) in CAD models, we propose a dual-resolution geometry extraction strategy that explicitly separates global shape from local detail.

\paragraph{Extracting face information}
For each face, to capture both its local and global geometric information, we perform UV parameterization~\cite{jayaraman2021uv} and uniformly sample two different resolution grids ($3\times 3$ for a global semantic guidance and $13\times 13$ for fine-grained local details).
We extract the coordinates (3D), normal (3D), and trimming indicator (1D) of each grid point to form a 7D feature as the face geometric information.
The face attribute for each grid point is a 16D vector that contains face type (6D), area (1D), centroid (3D), and bounding box (6D).

\paragraph{Extracting edge information}
Similar to the node feature construction, we uniformly sample 13 points on each BRep edge to extract the information of the edges.
The edge geometric information is a 12D vector, containing coordinate (3D), tangent (3D), and the normals of the two incident faces (6D), while the attribute is a 9D vector with edge type (5D), length (1D), and convexity (3D).

% \subsection{Pre-training}
% \label{sec:pre-training}
% \begin{figure*}[!t]
%   \centering
%   \includegraphics[width=\textwidth]{figs/pretraining.pdf}
%   \caption{Overview of Pretraining}
%   \label{fig:pretraining}
% \end{figure*}

\begin{figure}[t]
  \centering
  \includegraphics[width=0.95\linewidth]{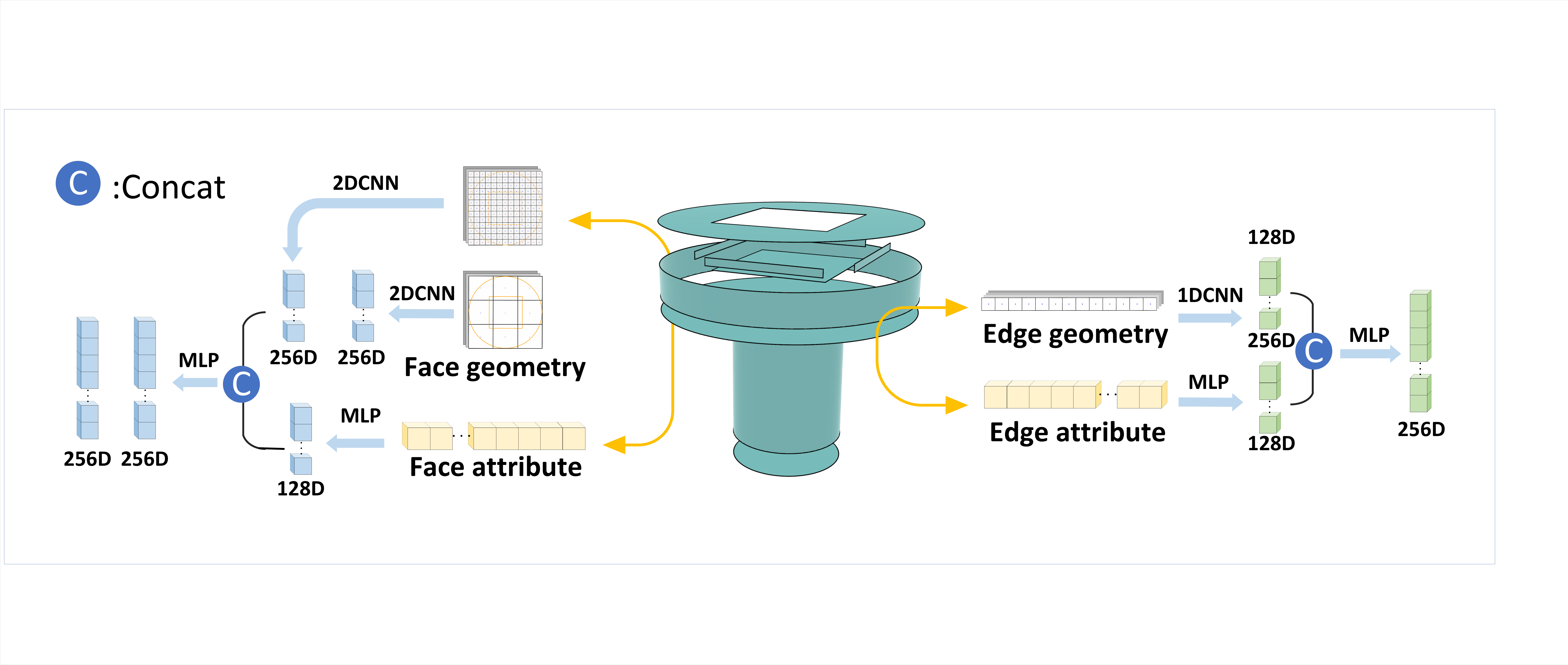}
    \vspace{-3mm}
  \caption{
  % For each face of a BRep model, we uniformly sample two different resolution grids on the UV domain for the geometry embeddings by a 2D CNN.
  % %
  % An MLP encodes the face attribute.
  % %
  % Then, the face geometry and attribute information are concatenated and processed by an MLP to generate face features.
  % %
  % Similarly, the edge geometry and attribute information are encoded by a 1D CNN and an MLP, respectively, and are concatenated as the input of an MLP to produce the edge feature.
  Our BRep encoder extracts and fuses geometric and attribute features into a unified gAAG representation via parallel CNN and MLP branches.
  }
  \label{fig:brep-encoding}
  \vspace{-2mm}
\end{figure}

\subsection{Pre-training} \label{sec:pre-training}

\subsubsection{Encoder}
Our encoder has two parts: (1) BRep encoder and (2) Graph encoder. %is a two-step process to first embed the raw BRep model into a unified latent graph which is then encoded into a and 

\paragraph{BRep encoder}
Raw BRep models consist of highly heterogeneous data, intricately intertwining continuous geometric with discrete topological information. 
To systematically process such heterogeneous data while strictly preserving its structural integrity, we project the BRep into a unified latent graph space via a gAAG~\cite{wu2024aagnet}. 
In this formulation, the gAAG nodes represent the extracted face features, while the graph edges represent the BRep topological edge features.

Specifically, our BRep encoder employs five parallel branches to independently embed the geometries and attributes: two 2D CNNs encode the face geometries, and an MLP encodes the attributes (Fig.~\ref{fig:brep-encoding}). 
The outputs of the two face geometry embeddings are both 256D vectors, while the attribute embedding is a 128D vector. 
We then concatenate each dual-resolution face geometric vector with its corresponding attribute vector, feeding them into another MLP to produce two decoupled 256D embeddings, denoted as $f^i_\text{low}$ and $f^i_\text{high}$, which serve as the node features for the gAAG.

Similarly, the geometries and attributes of BRep edges are extracted using a 1D CNN and an MLP. They are transferred to 256D and 128D vectors and then concatenated to generate a unified 256D embedding $e^i$ as the graph edge feature. 
Details are provided in the supplementary.

\paragraph{Graph encoder}
% A straightforward solution to process graph-structured data is to use GNNs.
% %
% However, the number of layers in GNN usually needs to be between 2-4 layers to achieve optimal results, as deep GNNs can lead to over-smoothing~\cite{zhang2021evaluating}. 
% %
% Therefore, GNN only aggregates local information while ignoring global information.
% %
% To overcome the issue, the graph encoder of our network is a GT, composed of two Transformer blocks and a GNN block (Fig.~\ref{fig:Graph_encoder}).

\begin{figure}[t]
  \centering
  % --- 左图: Graph Encoder ---
  \begin{minipage}[t]{0.4\linewidth}
    \centering
    % 图片宽度设为 minipage 的 100% (即全页宽的 48%)
    \includegraphics[width=0.99\linewidth]{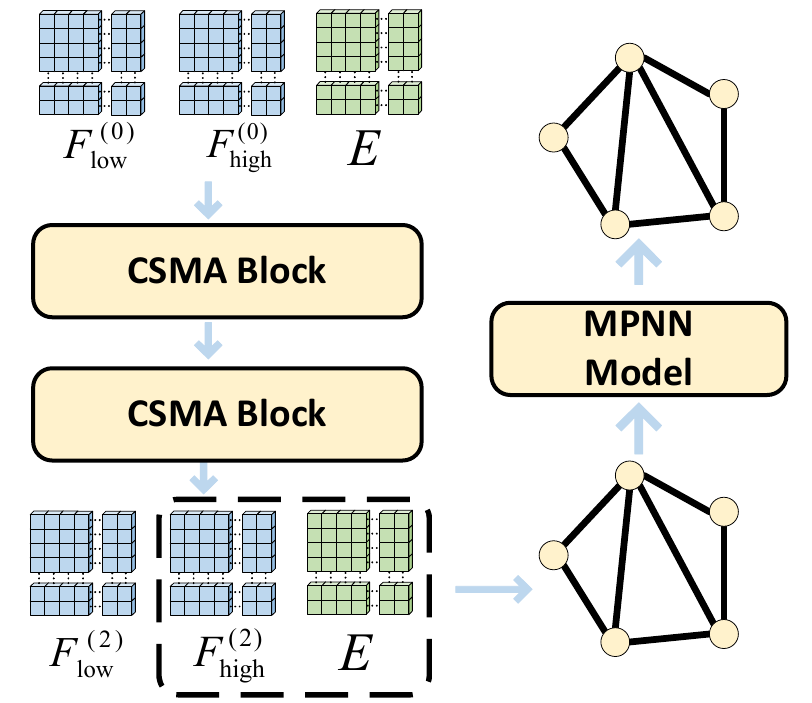}
    \vspace{-8mm}
    % \caption{
    %   Begining with the output of our BRep encoder ($F_{\text{low}}^{(0)} = F_{\text{low}}$, $F_{\text{medium}}^{(0)} = F_{\text{medium}}$, and $F_{\text{high}}^{(0)} = F_{\text{high}}$), the graph encoder first apply two Transformer blocks to extract the global information of BReps.
    %   Then, an MPNN block is introduced to update the face and edge features by utilizing neighborhood information.
    % }
    % \label{fig:Graph_encoder}
  \end{minipage}
  \hfill % 中间加弹簧撑开
  % --- 右图: Transformer ---
  \begin{minipage}[t]{0.55\linewidth}
    \centering
    \includegraphics[width=0.99\linewidth]{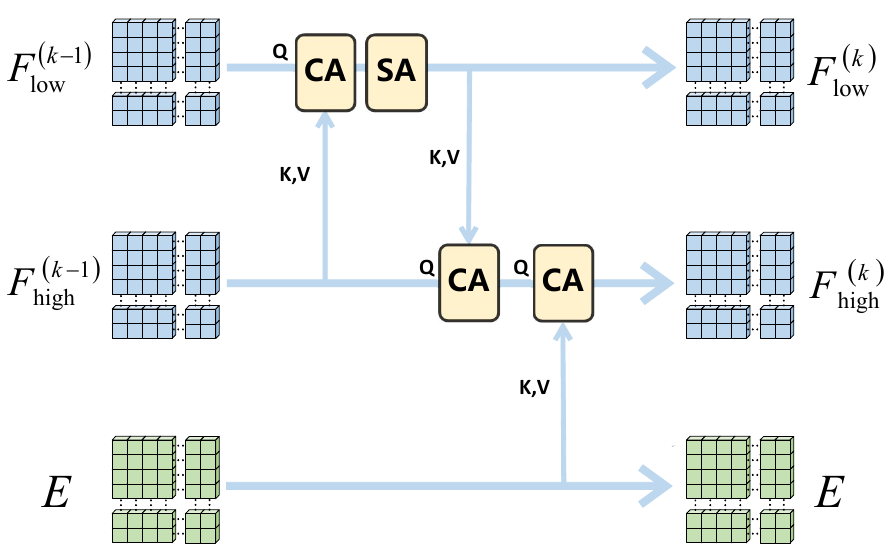}
    \vspace{-8mm}
    % \caption{
    %   In each CSMA block, the two scales are updated sequentially.
    %   First, $F_{\text{low}}^{(k-1)}$ cross-attends to $F_{\text{high}}^{(k-1)}$ and is then refined by a self-attention layer to yield $F_{\text{low}}^{(k)}$.
    %   Next, $F_{\text{high}}^{(k-1)}$, $F_{\text{low}}^{(k)}$, and the edge features $E$ are fused by two cross-attention layers to obtain $F_{\text{high}}^{(k)}$.
    %   CA and SA denote cross-attention and self-attention layers, respectively. Q, K, and V represent queries, keys, and values, respectively.
    % }
    % \label{fig:Transformer}
  \end{minipage}
  \caption{
      Left: The overall architecture of the Graph encoder, processing multi-resolution face features $F_\text{low}$, $F_\text{high}$, and edge features $E$. Right: The internal structure of the $k$-th CSMA block. CA and SA denote cross-attention and self-attention layers, respectively, while Q, K, and V represent queries, keys, and values.
    }
  \label{fig:encoder}
  \vspace{-2mm}
\end{figure}

Standard GNNs suffer from over-smoothing~\cite{zhang2021evaluating}, failing to capture long-range geometric dependencies. 
Simultaneously, directly applying standard Transformers to dense BRep samplings is inefficient, as the attention mechanism can be easily overwhelmed by redundant local details. 

To address this, our hierarchical graph Transformer encoder combines a cross-scale mutual attention (CSMA) module to capture global geometric structures and an MPNN~\cite{wu2024aagnet} to enforce local topological connections (Fig.~\ref{fig:encoder} - Left).
\begin{itemize}
    \item \textbf{Cross-Scale Mutual Attention.} 
To efficiently process features, we group node features by grid resolution into low-resolution features $F_{\text{low}}$ and high-resolution features $F_{\text{high}}$~\cite{Cho2025COD}, along with edge features $E$.
To make the network focus on key geometric changes rather than being distracted by large, flat surfaces, our CSMA block uses sparse, low-resolution features to query and extract essential details from dense, high-resolution features (Fig.~\ref{fig:encoder} - Right): 
\begin{equation}
F_{\text{low}}^{(k)} = \mathrm{SA}\!\Big(\mathrm{CA}\!\big(F_{\text{low}}^{(k-1)},\, F_{\text{high}}^{(k-1)}\big)\Big),
\end{equation}
\begin{equation}
F_{\text{high}}^{(k)} = \mathrm{CA}\!\Big(\mathrm{CA}\!\big(F_{\text{high}}^{(k-1)},\, F_{\text{low}}^{(k)}\big),\, E\Big), 
\end{equation}
where $\mathrm{CA}$ is multi-head cross-attention (queries from the first argument, keys/values from the second), and $\mathrm{SA}$ is self-attention. With $F_{\text{low}}^{(0)} = F_{\text{low}}$ and $F_{\text{high}}^{(0)} = F_{\text{high}}$, the updated $F_{\text{low}}^{(k)}$ aggregates a global context to guide the refinement of $F_{\text{high}}^{(k)}$ alongside explicit topological edge features $E$. 
We stack two such CSMA blocks to ensure the overall structure and local details are fully integrated. Crucially, this asymmetric bottleneck prevents the network from being overwhelmed by redundant flat surfaces and avoids the over-smoothing typical of GNNs. It ensures that crucial local details are accurately retained and integrated into the global context.
\item 
\textbf{Local Topological Message Passing.} 
Although the CSMA blocks capture the overall structure, BRep faces are geometrically connected by local edges. 
To explicitly model these precise neighborhood relationships, we pass the updated face features $F_\text{high}^{(2)}$ and edge features $E$ into an MPNN:
\begin{equation}
\big(F_{\text{latent}},\, E_{\text{latent}}\big)
= \mathrm{MPNN}\!\left(F_{\text{high}}^{(2)},\, E;\, \mathcal{G}_{\text{adj}}\right),
\end{equation}
where $\mathcal{G}_{\text{adj}}$ is the explicit adjacency graph. 
By placing this local MPNN after the global Transformer, our encoder effectively combines the overall 3D geometric structure with precise face-to-face connections. 
Detailed configurations are in the supplementary.
\end{itemize}

\subsubsection{Two-Stage Decoder}
Reconstructing heavily masked BRep models in a single step is highly challenging. Instead of directly predicting raw 3D geometries and attributes from the compressed latent space, we design a two-stage decoding strategy: a graph decoder followed by a BRep decoder.
\begin{enumerate}
    \item The graph decoder first reconstructs the intermediate node and edge features within the graph space. 
    \item Subsequently, the BRep Decoder maps these recovered features back to explicit geometries and attributes. 
\end{enumerate}
This two-stage breakdown not only eases the reconstruction but also allows the intermediate graph features to serve as additional supervision signals, thereby stabilizing the pre-training process.

\paragraph{Graph decoder}
To first recover the intermediate node and edge features from the masked context, the graph decoder employs an MPNN with an architecture symmetric to the graph encoder:
\begin{equation}
\big(F_{\text{recon}},\, E_{\text{recon}}\big)
= \mathrm{MPNN}\!\left(F_{\text{latent}},\, E_{\text{latent}};\, \mathcal{G}_{\text{adj}}\right),
\end{equation}
where $F_{\text{recon}}$ and $E_{\text{recon}}$ represent the reconstructed graph node and edge features, respectively. 
During training, these features are directly supervised by the features extracted from the unmasked BRep using the BRep encoder.

\paragraph{BRep decoder}
Subsequently, the BRep decoder utilizes four parallel branches to map these recovered graph features back to the explicit geometric and attribute details. 
For the continuous geometric structures, we employ FoldingNet~\cite{yang2018foldingnet} branches, which are naturally suited for deforming latent representations into 3D surfaces (7D face geometry) and curves (12D edge geometry). 
Simultaneously, two standard MLPs are used to regress the 16D face discrete attributes and 9D edge discrete attributes. 
Detailed layer-wise configurations are provided in the supplementary material.

\subsubsection{Masked Pre-training}
To learn robust BRep representations, we formulate a self-supervised masked autoencoding task. 
Crucially, instead of masking latent features~\cite{Yao2026BRepMAE}, we randomly mask the raw geometries and attributes of input faces and edges at a high ratio (e.g., 70\%). 
This strict input-level corruption forces the network to deduce missing structures from the surviving global context, preventing trivial local shortcuts.

Specifically, we reconstruct the edge features and only the high-resolution face features, along with their original geometric information. 
Since higher-resolution grids encapsulate richer geometric information, reconstructing them enables the model to accurately capture small geometric structures. %这里的fine-grained machining details是指什么？ //恢复13*13的grid能更准确地恢复细节特征，替换为了更直白的语言

Our objective includes five loss terms to provide comprehensive supervision across both latent features and explicit geometries. %一直没有理解什么是物理空间？前面也有类似的描述 //实际的三维空间，这里指的是不仅监督重建特征的信号(latent feature)，也监督实际的重建坐标结果(更改为了explicit geometries)
It includes a latent feature alignment loss $\mathcal{L}_\text{feat}$ to penalize the divergence between reconstructed graph features and the target features extracted from the original unmasked BRep. Crucially, a stop-gradient operation is applied to this target branch to prevent representation collapse and ensure the network learns meaningful semantics, alongside four explicit reconstruction losses ($\mathcal{L}_\text{geom}^\text{face}$, $\mathcal{L}_\text{attr}^\text{face}$, $\mathcal{L}_\text{geom}^\text{edge}$ and $\mathcal{L}_\text{attr}^\text{edge}$) for face/edge geometries and attributes. 
Detailed formulations are provided in the supplementary material.
We optimize the loss functions over all faces and edges to ensure our network accurately recovers the masked information.

\subsection{Fine-tuning}
\label{sec:domain-adaptive}
After pre-training, we attach a task-specific head behind the pre-trained encoder for downstream tasks.
The entire network is then fine-tuned by different loss functions with a small amount of labeled data from the target dataset.
Details for each downstream task are provided in the supplementary materials.

\section{Experiments}
\label{sec:Experments}
All experiments are implemented in PyTorch and executed on a server equipped with a single NVIDIA A800 GPU.
We first perform self-supervised pre-training, followed by task-specific evaluation across three downstream benchmarks: (1) machining feature recognition, (2) modeling segmentation, and (3) part classification.
Unless otherwise specified, all results are obtained under identical hardware configurations, and the best results are in bold for statistics comparisons.

% \subsection{Datasets}
% Our experiments use two groups of datasets: one for self-supervised pre-training and the other for downstream tasks. 

% \paragraph{Pre-training.}
% The pre-training dataset is constructed by merging the following datasets: MFInstSeg~\cite{wu2024aagnet} (62{,}495 models), MFCAD~\cite{cao2020graph} (15{,}488 models), MFCAD++~\cite{colligan2022hierarchical} (59{,}665 models), CADSynth~\cite{zhang2024brepmfr} (100{,}000 models), Fusion 360 Gallery Segmentation~\cite{lambourne2021brepnet} (35{,}858 models), and a 25{,}000-sample subset of SolidLetter~\cite{jayaraman2021uv}. 
% %
% This combined dataset comprises over 298{,}506 unique CAD models. 
% %
% Only BRep geometry and topology are used during pre-training, without any label supervision.

% \paragraph{Downstream tasks.}
% We evaluate our pre-trained encoder on the following labeled datasets:

% \begin{itemize}
%   \item \textbf{MFInstSeg} (62{,}495 models): per-face machining-feature labels; 25 classes.
%   \item \textbf{MFCAD++} (59{,}665 models): per-face machining-feature labels; 25 classes.
%   \item \textbf{CADSynth} (100{,}000 models): synthetic parts with per-face machining-feature labels; 25 classes.
%   \item \textbf{Fusion 360 Gallery Segmentation} (35{,}858 models): per-face modeling-operation labels; 8 classes.
%   \item \textbf{SolidLetter} (96{,}861 models): alphabetic CAD models with shape-level labels; 26 classes.
% \end{itemize}

\subsection{Datasets and Data Isolation Protocol}

To rigorously evaluate our method and prevent potential data leakage (i.e., data contamination), we enforce a strict data-isolation protocol across all experiments. 
Specifically, the CAD models are randomly split into training, validation, and testing subsets according to the task: a 70\%/15\%/15\% split is used for pre-training, model segmentation, and classification, while an 80\%/10\%/10\% split is adopted for machining feature recognition. 
Crucially, the testing sets are strictly held out and remain completely unseen during both the self-supervised pre-training and downstream fine-tuning phases, ensuring a fair evaluation.

\paragraph{Utilized benchmarks}
Our experiments utilize the following benchmarks:
\begin{itemize}
    \item MFInstSeg~\cite{wu2024aagnet} (62,495 models): per-face machining-feature labels; 25 classes.
    \item MFCAD++~\cite{colligan2022hierarchical} (59,665 models): per-face machining-feature labels; 25 classes.
    \item CADSynth~\cite{zhang2024brepmfr} (100,000 models): per-face machining-feature labels; 25 classes.
    \item Fusion 360 Gallery Segmentation~\cite{lambourne2021brepnet} (35,858 models): per-face modeling-operation labels; 8 classes.
    \item SolidLetter~\cite{jayaraman2021uv} (96,861 models): alphabetic CAD models with shape-level labels; 26 classes.
\end{itemize}

\paragraph{Pre-training corpus.} 
Following our strict isolation protocol, the pre-training dataset is constructed by merging only the training splits (70\%) of MFInstSeg, MFCAD++, CADSynth, Fusion 360 Gallery, and a 25,000-sample subset of SolidLetter. While the total collection of all datasets amounts to 283,018 models, this sanitized pre-training corpus comprises exactly 198,113 unique, unlabeled CAD models, ensuring absolute isolation from any downstream testing data.

\paragraph{Downstream adaptation.} 
For downstream tasks, we utilize the training and validation splits of the respective target datasets for fine-tuning, and report all final metrics exclusively on the strictly isolated testing splits.

\begin{figure}[t]
    \centering
    \includegraphics[width=0.99\linewidth]{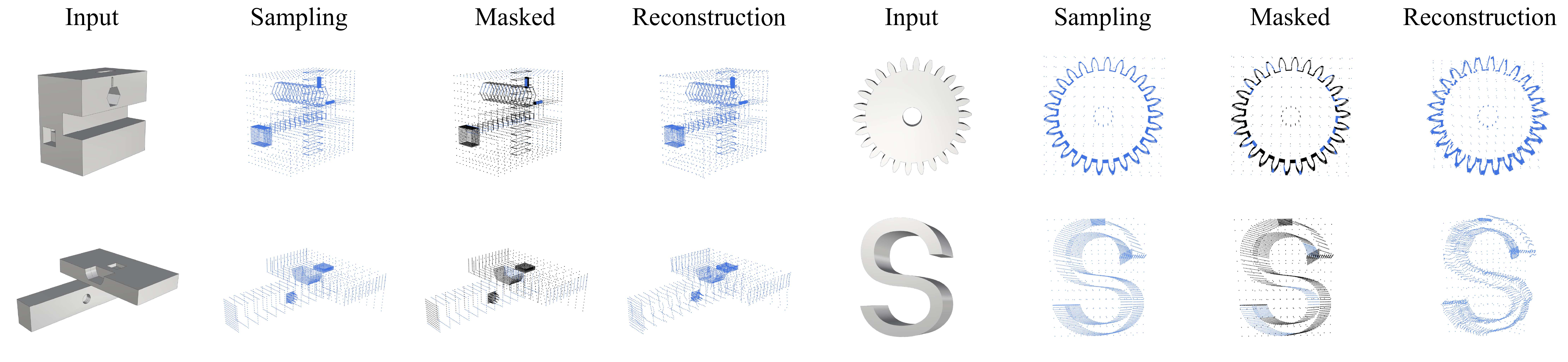}
    \vspace{-3mm}
    \caption{Four reconstruction examples. For each example, the sequence from left to right shows: (1) the input BRep model, (2) the full surface point cloud, (3) the masked input where black denotes masked faces and blue denotes unmasked faces, and (4) the reconstructed point cloud.}
    \label{fig:reconstruction-examples}
    \vspace{-2mm}
\end{figure}

\subsection{Pre-training}
In the pre-training stage, we use the AdamW optimizer with a batch size of 128 and an initial learning rate of $10^{-4}$. 
The learning rate follows a cosine decay schedule. 
Pre-training is conducted for 100 epochs from scratch, with 70\% of faces and edges randomly masked in each sample.
%
% Fig.~\ref{fig:pre-train-loss} shows the training loss over epochs, demonstrating stable convergence of the self-supervised objectives.
%
We show four examples of face-coordinate reconstruction in Fig.~\ref{fig:reconstruction-examples}.
The results show that our algorithm can correctly reconstruct the geometry of the masked BRep faces. 
Therefore, our algorithm does learn an effective BRep representation. 
%
% \tr{See supplementary for more reconstruction results.}
See supplementary for more reconstruction results.

\subsection{Comparisons}
We use face-level accuracy (Acc) and mean intersection over union (mIoU) to evaluate performance on machining feature recognition and modeling segmentation. For part classification, we report shape-level accuracy only.
%
% \tr{
% The formulas of Acc and mIoU are shown in the supplementary.
% }
The formulas of Acc and mIoU are shown in the supplementary.

\subsubsection{Machining feature recognition}

We evaluate our method on per-face machining-feature recognition (25 classes) using MFInstSeg, MFCAD++, and CADSynth. We compare against fully supervised baselines, AAGNet~\cite{wu2024aagnet} and the recent state-of-the-art BRepFormer~\cite{dai2025brepformer}, adopting their official implementations, alongside a recent self-supervised pre-training method, BRepMAE~\cite{Yao2026BRepMAE}. All datasets follow our strict 80/10/10 isolation protocol.

To assess few-shot performance, we fine-tune the network under varying supervision ratios (0.1\%--100\%). We use the AdamW optimizer for 200 epochs, setting learning rates to $10^{-4}$ for the task-specific head and $10^{-5}$ for the pre-trained encoder.

\begin{figure}[t]
    \centering
    \includegraphics[width=0.99\linewidth]{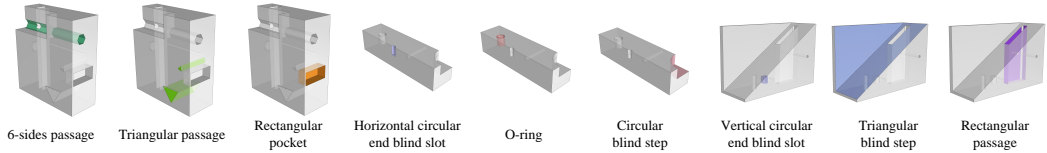}
 \vspace{-2mm}
    \caption{Machining feature recognition under the full supervised setting. We accurately predict various machining features, such as passages, pockets, slots, and steps.}
    \label{fig:Machining}

    \vspace{-4mm}
\end{figure}

As shown in Table~\ref{tab:all_seg_compact}, while the fully supervised BRepFormer demonstrates strong capacity under full supervision (100\%), training Transformers from scratch with extremely limited data (e.g., 0.1\%) is inherently challenging, leading to a drastic performance drop (e.g., falling to 33.43\% Acc and 7.47\% mIoU on MFInstSeg). In contrast, pre-training methods effectively alleviate this issue. Specifically, our method maintains a robust 88.75\% Acc and 66.38\% mIoU on MFInstSeg at the 0.1\% setting. It should be noted that while our method is strictly pre-trained on our 198K sanitized corpus, the BRepMAE results are reported from their original paper, which utilizes a significantly larger pre-training corpus of 250K models. Despite being pre-trained on substantially less unlabeled data, our method demonstrates superior data efficiency, consistently outperforming BRepMAE across most low-data regimes (e.g., achieving 96.19\% vs. 94.75\% Acc on MFInstSeg at 0.5\%, and 89.19\% vs. 86.56\% Acc on MFCAD++ at 0.1\%). This confirms that our hierarchical MAE pre-training extracts more transferable and robust representations. Furthermore, under full supervision, our method remains highly competitive with the fully supervised upper bound (e.g., 99.53\% vs. BRepFormer's 99.62\% Acc on MFInstSeg), proving that we achieve strong few-shot generalization without sacrificing the model's overall representational capacity. To further illustrate the effectiveness of our method, Figure~\ref{fig:Machining} shows representative results from the machining feature recognition task.

\begin{table}[t]
\centering
\caption{Comparison across MFInstSeg, MFCAD++, and CADSynth under varying supervision ratios. Accuracy (Acc, \%) and mIoU (\%) are reported. %Best results are in bold.
}
\label{tab:all_seg_compact}
%\scriptsize
\setlength{\tabcolsep}{4.2pt}
\renewcommand{\arraystretch}{1.1}
\scalebox{0.8}{
\begin{tabular}{c c lccccccc}
\toprule
& & \textbf{Method} & 0.1\% & 0.5\% & 1\% & 1.5\% & 2\% & 3\% & 100\% \\
\midrule
\multirow{6}{*}{\rotatebox{90}{\textbf{MFInstSeg}}}
& \multirow{3}{*}{\rotatebox{90}{Acc}} 
& AAGNet   & 46.49 & 76.62 & 85.19 & 90.30 & 92.31 & 92.37 & 99.13 \\
& & BRepFormer & 33.43 & 86.14 & 91.23 & 95.77 & 96.51 & 97.64 & \textbf{99.62} \\
& & BRepMAE  & \textbf{88.81}  & 94.75  & 96.99 & 97.44 & 97.66 & 98.20 & 99.57 \\
& & \textbf{Ours}  & 88.75 & \textbf{96.19} & \textbf{97.31} & \textbf{97.67} & \textbf{97.92} & \textbf{98.26} & 99.53 \\
\cmidrule(lr){2-10}
& \multirow{3}{*}{\rotatebox{90}{mIoU}} 

& AAGNet   & 34.01 & 67.51 & 77.50 & 84.64 & 86.76 & 87.32 & 98.36 \\
& & BRepFormer & 7.47 & 62.66 & 73.89 & 86.17 & 88.25 & 91.85 & \textbf{98.75} \\
& & BRepMAE  & \textbf{67.63}  & 83.38  & 90.39 & 91.33 & 91.93 & 93.59 & 98.60 \\
& & \textbf{Ours}     & 66.38 & \textbf{87.21} & \textbf{90.81} & \textbf{91.74} & \textbf{92.60} & \textbf{93.86} & 98.47 \\
\midrule
\multirow{6}{*}{\rotatebox{90}{\textbf{MFCAD++}}}
& \multirow{3}{*}{\rotatebox{90}{Acc}} 
& AAGNet   & 51.22 & 81.47 & 87.38 & 90.64 & 91.64 & 94.42 & 99.27 \\
& & BRepFormer  & 32.95 & 83.92 & 92.49 & 95.82 & 96.93 & 97.70 & \textbf{99.71} \\
& & BRepMAE  & 86.56  & 94.39  & 96.68 & 97.18 & 97.21 & 97.91 & 99.59 \\
& & \textbf{Ours}     & \textbf{89.19} & \textbf{95.94} & \textbf{97.34} & \textbf{97.60} & \textbf{97.97} & \textbf{98.36} & 99.61 \\
\cmidrule(lr){2-10}
& \multirow{3}{*}{\rotatebox{90}{mIoU}} 
& AAGNet   & 38.23 & 71.88 & 80.66 & 84.62 & 86.28 & 90.91 & 98.65 \\
& & BRepFormer  & 6.36  & 58.02  & 77.80 & 86.80 & 90.00 & 92.56 & \textbf{99.27} \\
& & BRepMAE  & 65.75  & 83.35  & 89.78 & 91.41 & 91.87 & 93.63 & 98.82 \\
& & \textbf{Ours}     & \textbf{67.29} & \textbf{86.26} & \textbf{90.97} & \textbf{92.12} & \textbf{93.25} & \textbf{94.57} & 99.03 \\
\midrule
\multirow{6}{*}{\rotatebox{90}{\textbf{CADSynth}}}
& \multirow{3}{*}{\rotatebox{90}{Acc}} 
& AAGNet   & 58.53 & 87.36 & 95.58 & 96.08 & 97.01 & 97.14 & 99.66 \\
& & BRepFormer  & 59.34 & 65.60 & 97.49 & 98.33 & 98.65 & 98.73 & \textbf{99.88} \\
& & BRepMAE  & 93.63  & 96.84  & 98.36 & 98.78 & 99.08 & 99.20 & 99.78 \\
& & \textbf{Ours}     & \textbf{93.82} & \textbf{98.54} & \textbf{98.91} & \textbf{99.25} & \textbf{99.29} & \textbf{99.42} & 99.86 \\
\cmidrule(lr){2-10}
& \multirow{3}{*}{\rotatebox{90}{mIoU}} 
& AAGNet   & 47.17 & 81.04 & 92.40 & 93.42 & 94.92 & 94.61 & 99.37 \\
& & BRepFormer  & 10.97  & 82.61  & 89.84 & 93.04 & 94.38 & 94.49 & \textbf{99.51} \\
& & BRepMAE  & 78.40  & 88.91  & 92.91 & 94.70 & 96.26 & 96.50 & 99.06 \\
& & \textbf{Ours}     & \textbf{79.29} & \textbf{93.72} & \textbf{95.53} & \textbf{96.85} & \textbf{97.03} & \textbf{97.55} & 99.43 \\
\bottomrule
\end{tabular} 
}
\end{table}

\begin{table}[t]
  \centering
  \caption{Performance comparison on Fusion 360 Gallery Segmentation under different levels of supervision. The $k$-shot setting denotes using $k$ labeled training samples in total. Accuracy (Acc, \%) and mIoU (\%) are reported. %Best results are in bold.
  }
  \label{tab:fusion360}
  %\small
  \renewcommand{\arraystretch}{1.1}
  \setlength{\tabcolsep}{5.0pt}
  \scalebox{0.8}{
  \begin{tabular}{lcccccc}
    \toprule
    \textbf{Model} & \multicolumn{2}{c}{\textbf{10-shot}} & \multicolumn{2}{c}{\textbf{20-shot}} & \multicolumn{2}{c}{\textbf{Full}} \\
    & Acc & mIoU & Acc & mIoU & Acc & mIoU \\
    \midrule
    BRep-BERT & 59.26 & 19.02 & 62.30 & 24.20 & 95.14 & 82.88 \\
    BRT & 54.98 & 13.54 & 56.69 & 14.99 & 94.48 & 79.23 \\
    BRepFormer & 50.24 & 12.31 &51.75 &13.38 &94.02 & 78.73\\
    Ours & \textbf{72.33} & \textbf{34.63} & \textbf{74.31} & \textbf{38.22} & \textbf{97.02} & \textbf{86.75} \\
    \bottomrule
  \end{tabular}
  }
\end{table}

\begin{figure}[t]
    \centering
    \includegraphics[width=0.99\linewidth]{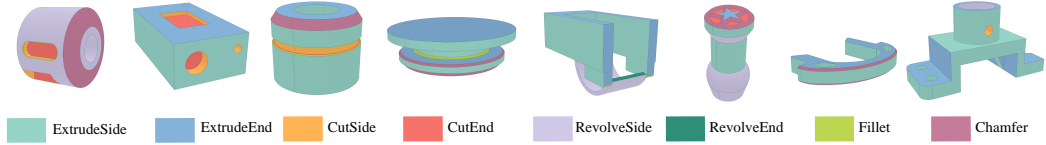}
    \vspace{0mm}
    \caption{Examples of modeling segmentation results on the Fusion 360 Gallery segmentation dataset under the full supervised setting. Our model accurately predicts per-face labels across 8 operation types, with distinct colors for each.}
    \label{fig:segmentation}
    \vspace{-4mm}
\end{figure}

\begin{figure}[t]
  \centering
  % --- 左图: Segmentation Compare ---
  \begin{minipage}[t]{0.42\linewidth}
    \centering
    \includegraphics[width=\linewidth]{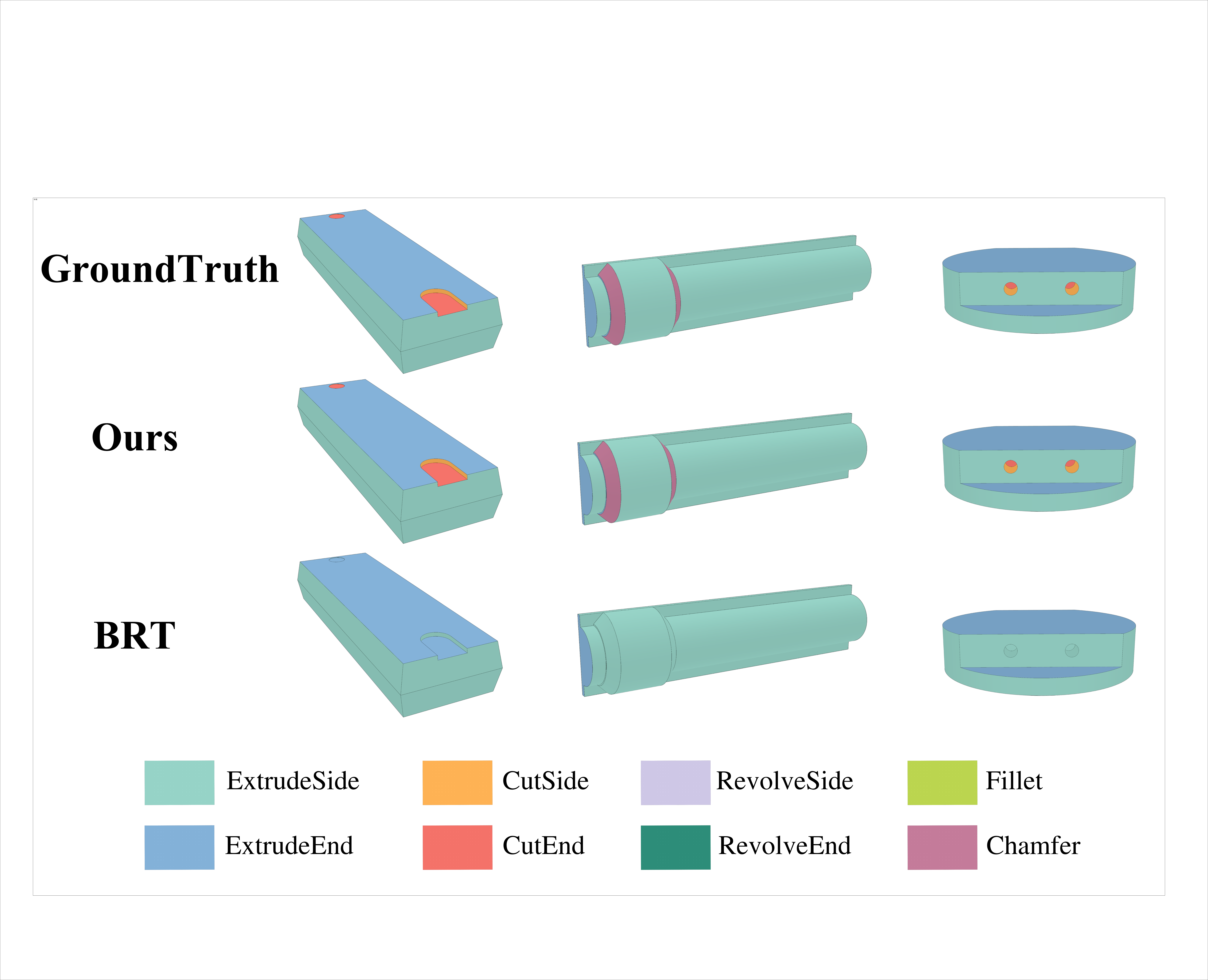}
    \vspace{-5mm}
    % \caption{Comparison with BRT under the 10-shot setting on the Fusion 360 Gallery segmentation dataset. Colors are consistent with Fig~\ref{fig:segmentation}. Our method predicts per-face labels more accurately.}
    % \label{fig:seg_compare}
  \end{minipage}
  \hfill % 中间撑开
  % --- 右图: Classification Compare ---
  \begin{minipage}[t]{0.55\linewidth}
    \centering
    \includegraphics[width=\linewidth]{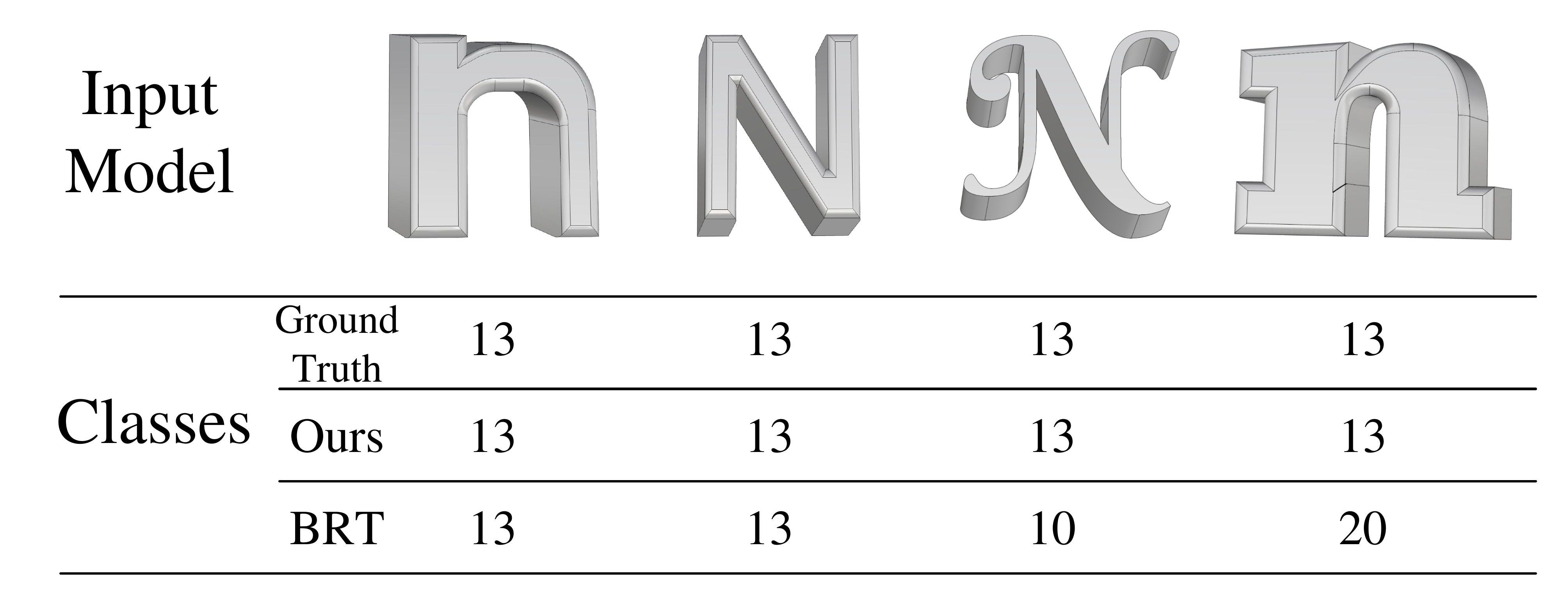}
    \vspace{-5mm}
    % \caption{10-shot comparison with BRT on the part classification task. Class IDs follow the dataset mapping: 13 = \texttt{n}, 10 = \texttt{k}, 20 = \texttt{u}. Our method classifies all four shapes as \texttt{n} and matches the ground truth; BRT is correct on the first two but predicts \texttt{k} for the third and \texttt{u} for the fourth.}
    % \label{fig:class_comp}
  \end{minipage}
  \caption{Performance comparison with BRT under the 10-shot setting. Left: The segmentation task, where our method predicts face labels more accurately. Right: The part classification task, where our method correctly identifies all shapes as \textbf{n} (class 13), while BRT misclassifies the last two.}
  \label{fig:comp_10shot}
\end{figure}

\subsubsection{Modeling segmentation}

We evaluate our pre-trained encoder on the Fusion 360 Gallery Segmentation dataset (8 classes). We compare against BRep-BERT~\cite{lou2023brep}, BRT~\cite{zou2025bringing}, and BRepFormer~\cite{dai2025brepformer} under 10-shot, 20-shot, and full-data settings, adopting their official implementations or reported results. The dataset is split into 70\% training, 15\% validation, and 15\% testing.

As shown in Table 2, our method achieves the highest performance across all settings. Under limited supervision (10-shot and 20-shot), fully supervised approaches like BRT and BRepFormer naturally experience performance drops since they are trained from scratch. Benefiting from the pre-trained representations, our model yields 72.33\% Acc and 34.63\% mIoU in the 10-shot scenario, demonstrating strong few-shot generalization and substantially exceeding the self-supervised baseline BRep-BERT (59.26\% Acc). Furthermore, while the fully supervised BRepFormer excels in machining feature recognition, its flat, single-level attention mechanism faces inherent challenges when handling the extreme scale variations of CAD surfaces. In contrast, our hierarchical architecture with the CSMA bottleneck and local message passing seamlessly integrates global geometric contexts with crucial local details. Consequently, our method not only excels in few-shot scenarios but also maintains the highest accuracy (97.02\%) and mIoU (86.75\%) under full supervision (See Fig.\ref{fig:segmentation}).

% \begin{figure}[t]
%   \centering
%   \includegraphics[width=0.7\linewidth]{figs/class_single.pdf}
%   \vspace{-4mm}
%   \caption{Part classification results on the SolidLetter dataset under the full supervised setting. Each row shows CAD models predicted to belong to the same alphabet class by our method.}
%   \vspace{-6mm}
%   \label{fig:classification}
% \end{figure}

\subsubsection{Part classification}
We evaluate our pre-trained encoder on the part classification task using the SolidLetter dataset, which consists of CAD models labeled across 26 alphabetic classes. We compare our method with strong baselines: BRep-BERT~\cite{lou2023brep}, BRT~\cite{zou2025bringing}, and BRepFormer~\cite{dai2025brepformer}. For BRep-BERT, we report results from the original paper, as the publicly available code is currently unavailable. For BRT and BRepFormer, we adopt their official implementations and follow the recommended hyperparameter settings.

To assess performance under different levels of supervision, we conduct experiments in three settings: 10-shot, 20-shot, and full-data. In the few-shot settings, 10 or 20 labeled examples are sampled per class for training. The dataset is split in a class-balanced manner: for each alphabet class, 70\% of the models are used for training, 15\% for validation, and 15\% for testing. All models are trained using the same experimental setup as described previously.

Our method achieves state-of-the-art performance on the SolidLetter benchmark across all supervision levels. Under 10-shot and 20-shot settings, our model achieves 82.51\% and 89.58\% accuracy, respectively, outperforming the self-supervised baseline BRep-BERT (68.71\% and 75.92\%) as well as fully supervised approaches like BRepFormer (76.53\% and 84.25\%) and BRT (54.84\% and 62.69\%) by substantial margins (See Fig~\ref{fig:comp_10shot} Right). Notably, even with very limited labeled data, our method maintains high classification accuracy, demonstrating strong generalization and data efficiency. Under full supervision, our method continues to yield the best performance, achieving 97.91\% accuracy, slightly surpassing BRep-BERT (97.76\%), BRepFormer (97.59\%), and BRT (97.35\%).

We further evaluate the cross-dataset generalization of our model by removing SolidLetter samples from the pre-training corpus (denoted as Ours*). As shown in Table~\ref{tab:solidletter}, even without exposure to the target dataset during pre-training, our method achieves 81.83\% and 88.47\% accuracy in the 10-shot and 20-shot settings, respectively, substantially outperforming all baselines, including the strong BRepFormer. Under full supervision, Ours* attains 97.78\%, closely matching the best results. These findings confirm the robustness and transferability of our pre-trained encoder across different geometric domains.

\begin{table}[t]
  \centering
  \caption{Comparison on SolidLetter classification under different levels of supervision. The $k$-shot setting denotes using $k$ labeled training samples for each class.  “Ours*” denotes our model pre-trained without SolidLetter data. Accuracy (Acc, \%) is reported. %Best results are in bold.
  }
  \label{tab:solidletter}
  %\small
  \setlength{\tabcolsep}{8.0pt} 
 \scalebox{0.8}{
  \begin{tabular}{lccc}
    \toprule
    \textbf{Model} & \textbf{10-shot Acc} & \textbf{20-shot Acc} & \textbf{Full Acc} \\
    \midrule
    BRep-BERT & 68.71 & 75.92 & 97.76 \\
    BRT       & 54.84 & 62.69 & 97.35 \\
    BRepFormer & 76.53 &84.25&97.59\\
    Ours* & 81.83 & 88.47 & 97.78 \\
    Ours      & \textbf{82.51} & \textbf{89.58} & \textbf{97.91} \\
    \bottomrule
  \end{tabular}
  }
  \vspace{-2mm}
\end{table}

\subsection{Ablation study}
\begin{table}[t]
\centering
\caption{Ablation study on MFInstSeg under various supervision ratios. The core architectural and pre-training designs are evaluated.}
\label{tab:ablation}
%\small
\setlength{\tabcolsep}{1.70pt}
\renewcommand{\arraystretch}{1.0}
\scalebox{0.8}{
\begin{tabular}{lccccccc}
\toprule
\textbf{Configuration} & \textbf{0.1\%} & \textbf{0.5\%} & \textbf{1\%} & \textbf{1.5\%} & \textbf{2\%} & \textbf{3\%} & \textbf{100\%} \\
\midrule

Single Res. ($13\times13$ only)  & 85.83 & 93.14 & 95.19 & 95.88 & 97.16 & 97.42 & 99.27 \\
Triple Res. (w/ $7\times7$)      & 86.41 & 94.46 & 96.40 & 97.04 & 97.38 & 97.78 & 99.42 \\
w/o Masking (0\% Mask)           & 66.41 & 84.19 & 92.99 & 94.20 & 95.33 & 95.96 & 99.43 \\
MPNN first                       & 76.39 & 92.10 & 94.95 & 95.62 & 96.12 & 96.41 & 99.06 \\
Mask gAAG        & 78.63 & 92.98 & 94.36 & 95.11 & 95.75 & 96.48 & 99.16 \\
Frozen encoder                   & 82.01 & 92.71 & 94.70 & 95.74 & 96.15 & 96.56 & 98.64 \\
\textbf{Ours (Default)}          & \textbf{88.75} & \textbf{96.19} & \textbf{97.31} & \textbf{97.67} & \textbf{97.92} & \textbf{98.26} & \textbf{99.53} \\
\bottomrule
\end{tabular}
}
\vspace{-4mm}
\end{table}

To isolate the contributions of our core designs, we conduct ablation studies on MFInstSeg (Table~\ref{tab:ablation}) across four critical dimensions. Additional ablations are detailed in the supplementary material.

\textbf{1. Dual-Resolution Information Bottleneck.}
We compare our dual-resolution ($3\times3$ and $13\times13$) CSMA block against \textit{Single Res.} ($13\times13$ only) and \textit{Triple Res.} (adding $7\times7$). \textit{Single Res.} consistently lags behind our method (e.g., 95.19\% vs. 97.31\% at 1\% data), indicating that lacking a sparse global anchor limits representation capability. Furthermore, \textit{Triple Res.} slightly underperforms our approach (e.g., 99.42\% vs. 99.53\% at 100\%), demonstrating that an intermediate resolution introduces mild attention dilution. Thus, our extreme coarse-to-fine bottleneck provides the optimal balance for feature compression.

\textbf{2. Necessity of Masked Modeling.}
Pre-training with simple autoencoding (0\% mask ratio, \textit{w/o Masking}) causes a drastic performance collapse in low-data regimes, dropping to 66.41\% at 0.1\% supervision. This $>$20\% gap confirms that high-ratio masking is essential, forcing the network to infer missing geometries from context rather than simply memorizing local features.

\textbf{3. Architectural Routing (Global to Local).}
Reversing our architecture to apply the MPNN before the Transformer (\textit{MPNN first}) significantly degrades performance (e.g., 76.39\% vs. 88.75\% at 0.1\%). This proves that establishing a global, coarse-to-fine geometric context before local topological message passing serves as a superior inductive bias for BRep data.

\textbf{4. Representation Transferability.}
When strictly freezing the pre-trained encoder and only updating the downstream MLP head (\textit{Frozen encoder}), the model still achieves a remarkable 82.01\% accuracy at 0.1\% supervision. This strong performance with frozen representations demonstrates that our pre-training framework extracts high-quality, task-agnostic geometric features.

\textbf{5. Masking strategies.}
Instead of masking inputs, we randomly mask the gAAG features after the BRep encoder. This consistently underperforms, confirming that input-level masking better guides pre-training.

\section{Conclusion} \label{sec:conclusion}
We propose the Masked BRep Autoencoder, a novel self-supervised framework for CAD representation learning. Our method effectively captures both the geometric and topological structures of BRep models by directly reconstructing the raw geometries and attributes of masked faces and edges, establishing a rigorous information bottleneck. By featuring a hierarchical architecture that seamlessly integrates global geometric contexts via the cross-scale mutual attention (CSMA) bottleneck and preserves crucial local details through explicit local message passing, our encoder learns rich, transferable representations.

Extensive experiments on machining feature recognition, modeling segmentation, and part classification benchmarks demonstrate the effectiveness and versatility of our approach. Notably, our method achieves strong performance with very limited supervision and remains competitive under full-data settings. These results highlight the practicality and generalizability of our framework for a wide range of downstream tasks. In the future, as larger datasets become available, our framework has the potential to scale to broader applications, such as BRep generative models.

\paragraph{Limitations}
Although our exhaustive dual-resolution sampling is crucial for capturing fine-grained details, it inherently increases the computational and storage overhead during data preprocessing. Consequently, scaling this face-level representation to massive industrial assemblies introduces significant memory bottlenecks, restricting the maximum sequence length during training.

%\clearpage  % TODO FINAL: This \clearpage needs to be removed from both review and camera-ready versions.

%\section*{Acknowledgements}
%Please insert your acknowledgments here.

% ---- Bibliography ----
%
% BibTeX users should specify bibliography style 'splncs04'.
% References will then be sorted and formatted in the correct style.
%

\bibliography{main}

\appendix
% \clearpage
% \setcounter{page}{1}
% \setcounter{section}{0}
% \setcounter{table}{0}
% \renewcommand{\thetable}{S\arabic{table}}
% \setcounter{figure}{0}
% \renewcommand{\thefigure}{S\arabic{figure}}
% % \title{Supplementary Material for: Masked BRep Autoencoder via Hierarchical Graph Transformer} 
% \maketitle
% \renewcommand{\thesection}{\Alph{section}}
% \renewcommand{\thesubsection}{\thesection.\arabic{subsection}}
% \renewcommand{\theequation}{S\arabic{equation}}

\section{BRep Encoder Details}
\label{sec:brep-encoder-supp}
\subsection{Input attribute}
The set of attribute information defined for each face is as follows:
\begin{itemize}
    \item \textbf{Type (6D):} One-hot vector indicating the surface type 
          (plane, cylinder, cone, sphere, torus, or NURBS).
    \item \textbf{BBox (6D):} The (x, y, z) coordinates of the two diagonal 
          vertices of the axis-aligned bounding box (AABB).
    \item \textbf{Centroid (3D):} The (x, y, z) coordinates of the face's centroid.
    \item \textbf{Area (1D):} The total area of the face.
\end{itemize}

Similarly, the set of attribute information defined for each edge is:
\begin{itemize}
    \item \textbf{Type (5D):} One-hot vector indicating the curve type 
          (Line, Circle, Ellipse, BSplineCurve, or Other).
    \item \textbf{Length (1D):} The total length of the edge.
    \item \textbf{Convexity (3D):} One-hot vector indicating the edge's convexity 
          (concave, convex, or smooth).
\end{itemize}

\subsection{Architecture of BRep encoder}
Our BRep encoder consists of 7 modules in total:
\begin{itemize}
    \item \textbf{Two 2D CNNs} for face geometry (one for each resolution). 
      These are separate instances that all use the same architecture (Table \ref{tab:2dcnn}).
    \item \textbf{One 1D CNN} for edge geometry (Table \ref{tab:1dcnn}).
    \item \textbf{One MLP} for face attribute (Table \ref{tab:face_attr_mlp}).
    \item \textbf{One MLP} for edge attribute (Table \ref{tab:edge_attr_mlp}).
    \item \textbf{One MLP} for face feature fusion (Table \ref{tab:fusion_mlp}).
    \item \textbf{One MLP} for edge feature fusion (Table \ref{tab:fusion_mlp}).
\end{itemize}

\begin{table}[h]
  \centering
  \caption{
    Architecture of the 2DCNN for face geometry.
    The same architecture is applied to process two face sampling resolutions. 
    The input grid size is generally denoted as $N \times N$, 
    where $N \in \{3, 13\}$.
    The convolution kernel size is $k$, padding is $p$, 
    and groups for GroupNorm are $G$.
  }
  \label{tab:2dcnn}
  \begin{tabular}{@{}lll@{}}
    \toprule
    Operator & Input Shape & Output Shape \\
    \midrule
    % Input (B, 7, N, N)
    Conv2D, $k=3, p=1$ & $7 \times N \times N$ & $64 \times N \times N$ \\
    GroupNorm, $G=8$ & $64 \times N \times N$ & $64 \times N \times N$ \\
    ReLU & $64 \times N \times N$ & $64 \times N \times N$ \\
    \midrule
    Conv2D, $k=3, p=1$ & $64 \times N \times N$ & $128 \times N \times N$ \\
    GroupNorm, $G=16$ & $128 \times N \times N$ & $128 \times N \times N$ \\
    ReLU & $128 \times N \times N$ & $128 \times N \times N$ \\
    \midrule
    Conv2D, $k=3, p=1$ & $128 \times N \times N$ & $256 \times N \times N$ \\
    GroupNorm, $G=32$ & $256 \times N \times N$ & $256 \times N \times N$ \\
    ReLU & $256 \times N \times N$ & $256 \times N \times N$ \\
    \midrule
    % --- This row clearly shows the resolution change ---
    AdaptiveAvgPool2D & $256 \times N \times N$ & $256 \times 1 \times 1$ \\
    Flatten & $256 \times 1 \times 1$ & $256$ \\
    \bottomrule
  \end{tabular}
\end{table}

\begin{table}[t]
  \centering
  \caption{
    Architecture of the 1DCNN for edge geometry.
    The input is a sequence of 13 points sampled along the edge, 
    each with a 12D vector.
    The convolution kernel size is $k$, the padding is $p$, and the groups for GroupNorm are $G$.
  }
  \label{tab:1dcnn}
  % \small
  % \renewcommand{\arraystretch}{1.1}
  \setlength{\tabcolsep}{10.0pt}  
  \begin{tabular}{@{}lcc@{}}
    \toprule
    Operator & Input Shape & Output Shape \\
    \midrule
    Conv1D, $k=3, p=1$ & $12 \times 13$ & $64 \times 13$ \\
    GroupNorm, $G=8$ & $64 \times 13$ & $64 \times 13$ \\
    ReLU & $64 \times 13$ & $64 \times 13$ \\
    \midrule
    Conv1D, $k=3, p=1$ & $64 \times 13$ & $256 \times 13$ \\
    GroupNorm, $G=32$ & $256 \times 13$ & $256 \times 13$ \\
    ReLU & $256 \times 13$ & $256 \times 13$ \\
    \midrule
    AdaptiveAvgPool1D & $256 \times 13$ & $256 \times 1$ \\
    Flatten & $256 \times 1$ & $256$ \\
    \bottomrule
  \end{tabular}
\end{table}

\begin{table}[t]
  \centering
  \caption{
    Architecture of the MLP for face attribute.
  }
  \label{tab:face_attr_mlp}
  \setlength{\tabcolsep}{21.0pt}  
  \begin{tabular}{@{}lcc@{}}
    \toprule
    Operator & Input Shape & Output Shape \\
    \midrule
    Linear & 16 & 128 \\
    LayerNorm & 128 & 128 \\
    GELU & 128 & 128 \\
    Linear & 128 & 128 \\
    \bottomrule
  \end{tabular}
\end{table}

\begin{table}[t]
  \centering
  \caption{
    Architecture of the MLP for edge attribute.
  }
  \label{tab:edge_attr_mlp}
  \setlength{\tabcolsep}{21.0pt}  
  \begin{tabular}{@{}lcc@{}}
    \toprule
    Operator & Input Shape & Output Shape \\
    \midrule
    Linear & 9 & 128 \\
    LayerNorm & 128 & 128 \\
    GELU & 128 & 128 \\
    Linear & 128 & 128 \\
    \bottomrule
  \end{tabular}
\end{table}

\begin{table}[t]
  \centering
  \caption{
    Architecture of the MLP for feature fusion.
    This network is used to fuse the geometric features (256D) 
    and attribute features (128D) into a final 256D vector.
    The input shape is the concatenated dimension ($256+128=384$).
  }
  \label{tab:fusion_mlp}
  \setlength{\tabcolsep}{21.0pt}  
  \begin{tabular}{@{}lcc@{}}
    \toprule
    Operator & Input Shape & Output Shape \\
    \midrule
    Linear & 384 & 256 \\
    LayerNorm & 256 & 256 \\
    GELU & 256 & 256 \\
    Linear & 256 & 256 \\
    \bottomrule
  \end{tabular}
\end{table}

\section{Hierarchical Graph Transformer Details}
\subsection{Architecture of Cross-Scale Mutual Attention (CSMA) Block}
The CSMA block is built from two distinct components:

\begin{itemize}
    \item \textbf{Self-Attention Block (SA):} Our SA block is a standard Transformer encoder layer. It uses a Post-LayerNorm design and consists of one Multi-Head Attention module followed by a Feed-Forward Network (FFN).
    \item \textbf{Cross-Attention Block (CA):} Our CA block is a simplified module designed only for feature aggregation. It only consists of one Multi-Head Attention module.
\end{itemize}

The key hyperparameters used to configure both SA and CA blocks are shared and detailed in Table \ref{tab:gt_hyperparams}.

\begin{table}[h]
  \centering
  \caption{
    Hyperparameters for our Transformer blocks.
  }
  \label{tab:gt_hyperparams}
  \setlength{\tabcolsep}{48.0pt}  
  \begin{tabular}{@{}lc@{}}
    \toprule
    Hyperparameter & Value \\
    \midrule
    Embedding Dimension  & 256 \\
    Number of Attention Heads & 8 \\
    FFN Hidden Dimension  & 1024 \\
    FFN Activation & GELU \\
    Dropout Rate & 0.3 \\
    \bottomrule
  \end{tabular}
\end{table}

\subsection{Architecture of MPNN Block}

We adopt a Message Passing Neural Network (MPNN) block to propagate information across the gAAG topology. The update consists of three steps: edge update, message aggregation, and node update.

Denoting $f_i^{(l)}$ as the feature of node $i$ at layer $l$, and $e_{ji}^{(l)}$ as the feature of the edge from node $j$ to $i$, the update is defined as follows:
\begin{itemize}
  \item \textbf{Edge update:} The edge feature is updated by combining features from its incident nodes and current edge embedding:
  \begin{equation}
       e_{ji}^{(l+1)} = \mathrm{MLP}\left(f_j^{(l)} \, \| \, f_i^{(l)} \, \| \, e_{ji}^{(l)}\right),\ \ l = 0, 1, \dots, L{-}1
  \end{equation}
  where $\|$ denotes vector concatenation, and $L$ is the total number of MPNN layers. 
  In our implementation, we use $L = 2$.
  The input dimension of the MLP is $256 \times 3$, followed by a hidden layer of dimension 256 with ReLU activation, and an output layer of dimension 256.

  \item \textbf{Message aggregation:} Node $i$ aggregates messages from its neighbors using a graph convolutional network GAT :
  \begin{equation}
      \tilde{f}_i^{(l+1)} = \mathrm{GAT}\left(f_i^{(l)}, \left\{(f_j^{(l)}, e_{ji}^{(l+1)}) \mid j \in \mathcal{N}(i)\right\} \right),
  \end{equation}
  where $\mathcal{N}(i)$ is the set of indices for all neighbor nodes of the $i$th node.

  \item \textbf{Node update:} The aggregated message is passed through a two-layer feed-forward network (FFN), followed by residual addition and layer normalization:
  \[
  f_i^{(l+1)} = \mathrm{LayerNorm}\left(f_i^{(l)} + \mathrm{FFN}\left(\tilde{f}_i^{(l+1)}\right)\right).
  \]
   The FFN consists of two linear layers with GELU activation. The input and output dimensions are both 256, while the hidden layer has a dimension of 512.
\end{itemize}

\section{BRep Decoder Details}
The BRep decoder consists of four parallel branches to reconstruct the face and edge information from latent graph features. Specifically:

\begin{itemize}
    \item \textbf{Face Geometry Branch:} This branch employs a two-stage FoldingNet decoder to reconstruct the 7D geometric features for each face sampling point, including 3D coordinates, 3D normals, and a 1D trimming indicator. The decoder follows a two-stage architecture composed of MLPs. In the first stage, a 2D grid point \((u, v)\) is concatenated with the 256D reconstructed face feature and passed through a three-layer MLP to generate a 64D intermediate feature. In the second stage, this 64D feature is concatenated with the original face feature and further processed by another three-layer MLP to produce the final 7D output. The detailed architecture is summarized in Table~\ref{tab:face_geom_decoder}.

    \item \textbf{Face Attribute Branch:} This branch is a single MLP that regresses the 16D face attributes, including face type, area, centroid, and bounding box information. See Table~\ref{tab:de_face_attr} for details.

    \item \textbf{Edge Geometry Branch:} Similar to the face geometry branch, this decoder employs a two-stage FoldingNet structure to reconstruct the 12D geometric features of each edge sampling point, including 3D coordinates, tangent vectors, and incident face normals. The first stage takes the edge feature concatenated with a 1D curve parameter \( u \), producing a 64D intermediate representation. The second stage refines this by concatenating it again with the edge feature and decoding it into a final 12D output. See Table~\ref{tab:edge_geom_decoder} for details.

    \item \textbf{Edge Attribute Branch:} A lightweight MLP regresses the 9D edge attribute, including edge type, length, and convexity. See Table~\ref{tab:de_edge_attr} for details.
\end{itemize}

\begin{table}[t]
\centering
\caption{Architecture of the FoldingNet decoder for face geometry.}
\label{tab:face_geom_decoder}
\setlength{\tabcolsep}{16.0pt}  
\begin{tabular}{lcc}
\toprule
Operator & Input Shape & Output Shape \\
\midrule
Linear & $256+2$ & 512 \\
ReLU & 512 & 512 \\
Linear & 512 & 512 \\
ReLU & 512 & 512 \\
Linear & 512 & 64 \\
\midrule
Linear & $256+64$ & 512 \\
ReLU & 512 & 512 \\
Linear & 512 & 512 \\
ReLU & 512 & 512 \\
Linear & 512 & 7 \\
\bottomrule
\end{tabular}
\end{table}

\begin{table}[h]
  \centering
  \caption{
  Architecture of the MLP decoder for face attribute information.
  }
  \label{tab:de_face_attr}
  \setlength{\tabcolsep}{21.0pt}  
  \begin{tabular}{@{}lcc@{}}
    \toprule
    Operator & Input Shape & Output Shape \\
    \midrule
    Linear & 256 & 256 \\
    LayerNorm & 256 & 256 \\
    GELU & 256 & 256 \\
    Linear & 256 & 16 \\
    \bottomrule
  \end{tabular}
\end{table}

\begin{table}[t]
\centering
\caption{Architecture of the FoldingNet decoder for edge geometry.}
\label{tab:edge_geom_decoder}
\setlength{\tabcolsep}{16.0pt}  
\begin{tabular}{lcc}
\toprule
Operator & Input Shape & Output Shape \\
\midrule
Linear & $256+1$ & 512 \\
ReLU & 512 & 512 \\
Linear & 512 & 512 \\
ReLU & 512 & 512 \\
Linear & 512 & 64 \\
\midrule
Linear & $256+64$ & 512 \\
ReLU & 512 & 512 \\
Linear & 512 & 512 \\
ReLU & 512 & 512 \\
Linear & 512 & 12 \\
\bottomrule
\end{tabular}
\end{table}

\begin{table}[h]
  \centering
  \caption{
  Architecture of the MLP decoder for edge attribute.
  }
  \label{tab:de_edge_attr}
  \setlength{\tabcolsep}{21.0pt} 
  \begin{tabular}{@{}lcc@{}}
    \toprule
    Operator & Input Shape & Output Shape \\
    \midrule
    Linear & 256 & 256 \\
    LayerNorm & 256 & 256 \\
    GELU & 256 & 256 \\
    Linear & 256 & 9 \\
    \bottomrule
  \end{tabular}
\end{table}

\section{Losses for pre-training}
We define our pre-training loss function as follows:
\begin{equation}
    \mathcal{L}
= \lambda_{1}\mathcal{L}_{\text{feat}}
+ \lambda_{2}\mathcal{L}_{\text{geom}}^{\text{face}}
+ \lambda_{3}\mathcal{L}_{\text{attr}}^{\text{face}}
+ \lambda_{4}\mathcal{L}_{\text{geom}}^{\text{edge}}
+ \lambda_{5}\mathcal{L}_{\text{attr}}^{\text{edge}},
\end{equation}
where $\mathcal{L}_{\text{feat}}$ is used to reconstruct the graph node and edge features obtained from the masked geometric information by our BRep encoder, and $\mathcal{L}_{\text{geom}}^{\text{face}}$, $\mathcal{L}_{\text{attr}}^{\text{face}}$, $\mathcal{L}_{\text{geom}}^{\text{edge}}$, and $\mathcal{L}_{\text{attr}}^{\text{edge}}$ measures the difference between the input BRep information and the reconstructed information.
In our experiments, we set $ \lambda_{1} = 1$, $ \lambda_{2} = 1$, $ \lambda_{3} = 0.3$, $ \lambda_{4} = 0.5$, and $ \lambda_{5} = 0.3$.

\paragraph{Feature term}
The feature term ensures that the face and edge features obtained through the graph decoder are consistent with the ground truth.
Therefore, we adopt the mean squared error (MSE) to measure the difference:
\begin{equation}
\begin{aligned}
\mathcal{L}_{\text{feat}}
= &\frac{1}{|\mathcal{M}_f|}\sum_{i\in\mathcal{M}_f}\!\big\|F_{\text{recon}, i}^{m}-F_{\text{high}, i}^t\big\|_2^2 +\\
&  \frac{1}{|\mathcal{M}_e|}\sum_{i\in\mathcal{M}_e}\!\big\|E_{\text{recon}, i}^{m}-E_{\text{high}, i}^t\big\|_2^2,
\end{aligned}
\end{equation}
where $\mathcal{M}_f$ and $\mathcal{M}_e$ are the sets of masked faces and edges, respectively, and $|\mathcal{M}_f|$ and $|\mathcal{M}_e|$ are the number of masked faces and edges, respectively.
$F_{\text{recon}, i}^{m}$ and $E_{\text{recon}, i}^{m}$ are the reconstructed face and edge features by our decoder.
$F_{\text{high}, i}^t$ and $E_{\text{high}, i}^t$ are the ground truth, which are generated by the BRep encoder with the original geometric information corresponding to the masked faces and edges as its inputs.

%Pretraining is supervised by a feature-level reconstruction term on the masked embeddings and by geometry-level objectives on grids and attributes. 
%Let $\mathcal{F}$ and $\mathcal{E}$ be the index sets of faces and edges, and let $\mathcal{M}_f\!\subseteq\!\mathcal{F}$ and $\mathcal{M}_e\!\subseteq\!\mathcal{E}$ denote the masked subsets. 
%Denote the ground-truth face/edge feature embeddings by $F\in\mathbb{R}^{|\mathcal{F}|\times d_f}$ and $E\in\mathbb{R}^{|\mathcal{E}|\times d_e}$, with reconstructions $F_{\text{recon}}$ and $E_{\text{recon}}$. 
%We use the mean squared error (MSE) on the masked entries:
%\begin{equation}
%\begin{aligned}
%\mathcal{L}_{\text{feat}}
%&= \frac{1}{|\mathcal{M}_f|}\sum_{i\in\mathcal{M}_f}\!\big\|F_{\text{recon}}[i]-%F[i]\big\|_2^2 \\
%&\quad + \frac{1}{|\mathcal{M}_e|}\sum_{j\in\mathcal{M}_e}\!\big\|E_{\text{recon}}[j]-E[j]\big\|_2^2 .
%\end{aligned}\tag{5}
%\end{equation}
\paragraph{Face geometry term}
The face geometric information of each sampling point is a 7D vector $(\mathbf{p}_{i},\,\mathbf{n}_{i},\,\tau_{i})$, where $\mathbf{p}_{i}$ is its coordinate, $\mathbf{n}_{i}$ is its face normal, and $\tau_{i}$ indicates whether the point is trimmed ($\tau_{i} = 0$ means it is trimmed and $\tau_{i} = 1$ means it is retained).
Considering that the components of face geometric information represent different physical meanings, we use different loss functions and weights to calculate them respectively:
\begin{equation}
\mathcal{L}_{\text{geom}}^{\text{face}} = \alpha_1\,\mathcal{L}_{\text{coord}}^{\text{face}}
+ \alpha_2\,\mathcal{L}_{\text{norm}}^{\text{face}}
+ \alpha_3\,\mathcal{L}_{\text{trim}}^{\text{face}},
\end{equation}
where
\begin{equation}
\mathcal{L}_{\text{coord}}^{\text{face}}=\frac{1}{|\mathcal{P}_f|}\sum_{i\in\mathcal{P}}
\big\|\widehat{\mathbf{p}}_{i}-\mathbf{p}_{i}\big\|_2^2,
\end{equation}
\begin{equation}
\mathcal{L}_{\text{norm}}^{\text{face}}=\frac{1}{|\mathcal{P}_f|}\sum_{i\in\mathcal{P}}
\big\|\widehat{\mathbf{n}}_{i}-\mathbf{n}_{i}\big\|_2^2,
\end{equation}
\begin{equation}
\mathcal{L}_{\text{trim}}^{\text{face}}=\frac{1}{|\mathcal{P}_f|}\sum_{i\in\mathcal{P}}
\ell_{\mathrm{bce}}\!\big(\widehat{\tau}_{i},\,\tau_{i}\big).
\end{equation}
$\mathcal{P}_f$ is the set of sampling points with the highest resolution on all BRep faces, and $|\mathcal{P}_f|$ is its number.
$\widehat{\mathbf{p}}_{i}$, $\widehat{\mathbf{n}}_{i}$, and $\widehat{\tau}_{i}$ are the predicted geometric information of these sampling points, respectively.
$\ell_{\mathrm{bce}}$ is the binary cross-entropy (BCE) function defined as follows:
\begin{equation}
    \ell_{\mathrm{bce}}(\hat{\tau},\tau)
= -\big(\tau\log \hat{\tau} + (1-\tau)\log(1-\hat{\tau})\big).
\end{equation}
In our experiments, we set $\alpha_1 = 1$, $\alpha_2 = 0.5$, and $\alpha_3 = 0.3$.

\paragraph{Face attribute term}
The face attribute information is recovered from the graph node feature by an MLP.
We use the following loss function to encourage the predicted face attribute information to be close to the ground truth value:
\begin{equation}
    \mathcal{L}_{\text{attr}}^{\text{face}}
= \frac{1}{|\mathcal{F}|}\sum_{i\in\mathcal{F}}
\big\|\widehat{a}^{\,\text{face}}_i-a^{\text{face}}_i\big\|_2^2,
\end{equation}
where $\mathcal{F}$ is the set of BRep faces, and $|\mathcal{F}|$ is the number of BRep faces.
$\widehat{a}^{\,\text{face}}_i$ and $a^{\text{face}}_i$ are the predicted and original face attribute vectors, respectively.

\paragraph{Edge geometry term}
Since the edge lacks trimming information, we use MSE as the loss function over the entire 12D edge information:
\begin{equation}
    \mathcal{L}_{\text{geom}}^{\text{edge}}
= \frac{1}{|\mathcal{P}_e|}\sum_{i\in\mathcal{P}_e}
\big\|\widehat{\mathbf{c}}_{i}-\mathbf{c}_i\big\|_2^2,
\end{equation}
where $\mathcal{P}_e$ is the set of sampling points on all BRep edges, and $|\mathcal{P}_e|$ is the number of edge sampling points.
$\widehat{\mathbf{c}}_{i}$ and $\mathbf{c}_i$ are the predicted and original edge geometric information, respectively.

\paragraph{Edge attribute term}
The same as the face attribute term, our edge attribute loss function is defined as follows:
\begin{equation}
    \mathcal{L}_{\text{attr}}^{\text{edge}}
= \frac{1}{|\mathcal{E}|}\sum_{i\in\mathcal{E}}
\big\|\widehat{a}^{\,\text{edge}}_i-a^{\text{edge}}_i\big\|_2^2
\end{equation}
where $\mathcal{E}$ is the set of BRep edges, and $|\mathcal{E}|$ is its number.
$\widehat{a}^{\,\text{edge}}_i$ and $a^{\text{edge}}_i$ are the recovered and corresponding 9D BRep edge attribute vectors, respectively.

\section{Fine-tuning Details}
Our pre-trained encoder is adapted for downstream tasks by attaching a small, task-specific head (see table~\ref{tab:task_head}) and fine-tuned on a small labeled dataset with task-specific losses.
\paragraph{Loss for machining feature recognition}
For the machining feature recognition task, the task-specific head takes each face's feature $f_\text{latent}^i = (F_\text{latent})_i$, the $i$-th row of $F_{\text{latent}}$, as input and outputs a vector $\mathbf{z}^{\text{mach}}_i\in\mathbb{R}^{C_{\text{mach}}}$.
The constant $C_{\text{mach}}$ is the number of machining features.
We use the following loss function to train our network:
\begin{equation}
    \mathcal{L}_{\text{mach}}=\frac{1}{N_f}\sum_{i=1}^{N_f}\ell_{\mathrm{ce}}(\mathbf{z}^{\text{mach}}_i,\mathbf{y}^{\text{mach}}_i),
\end{equation}
where $N_f$ is the number of faces in the BRep model and $\mathbf{y}^{\text{mach}}_i$ is a one-hot vector representing the ground truth.
$\ell_{\mathrm{ce}}$ is the multiclass cross-entropy function:
\begin{equation}
    \ell_{\mathrm{ce}}(\mathbf{z},\mathbf{y})=-\sum_{i=1}^{C} y_i\log \frac{\exp(z_i)}{\sum_{j=1}^{C}\exp(z_j)},
\end{equation}
where $C$ is the dimension of $\mathbf{z}$ and $\mathbf{y}$, indicating the number of classes.

\paragraph{Loss for modeling-segmentation}
The same as the machining feature recognition task, the input of our task-specific head is $f_\text{latent}^i$, and the output is $\mathbf{z}^{\text{seg}}_i$.
Our modeling-segmentation loss function is defined as follows:
\begin{equation}
    \mathcal{L}_{\text{seg}}=\frac{1}{N_f}\sum_{i=1}^{N_f}\ell_{\mathrm{ce}}(\mathbf{z}^{\text{seg}}_i,\mathbf{y}^{\text{seg}}_i),
\end{equation}
where $\mathbf{z}^{\text{seg}}_i\in\mathbb{R}^{C_{\text{seg}}}$ is the output of our network, $\mathbf{y}^{\text{seg}}_i$ is the ground truth and $C_{\text{seg}}$ is the number of segmentation classes.

\paragraph{Loss for classification}
Since the classification task involves classifying the entire BRep, we employ a global max pooling layer to aggregate the BRep face features $F_\text{latent}$ as the BRep feature $B_\text{latent}$:
\begin{equation}
    (B_{\text{latent}})_j=\max_{1\le i\le N_f} (F_{\text{latent}})_{i,j}.
\end{equation}
Then we input $B_\text{latent}$ into the task-specific head to generate a vector $\mathbf{z}_\text{cls}\in\mathbb{R}^{C_{\text{cls}}}$, where $C_{\text{cls}}$ is the number of CAD classes as output.
The loss function is as follows:
\begin{equation}
    \mathcal{L}_{\text{cls}}=\ell_{\mathrm{ce}}(\mathbf{z}_{\text{cls}},\mathbf{y}_{\text{cls}}),
\end{equation}
where $\mathbf{y}_{\text{cls}}$ is the ground truth one-hot vector.

\begin{table}[h]
  \centering
  \caption{
  Architecture of the downstream task head. 
It consists of three linear layers with GELU activations and dropout (rate = 0.3) applied after each hidden layer. 
$C$ denotes the number of output classes. The input is the encoded feature vector produced by our pre-trained encoder. 
For machining feature recognition and modeling segmentation, each face's feature is fed independently. 
For part classification, a global max pooling is applied over all face features to obtain the input.
  }
  \label{tab:task_head}
  \setlength{\tabcolsep}{23.0pt} 
  \begin{tabular}{@{}lcc@{}}
    \toprule
    Operator & Input Shape & Output Shape \\
    \midrule
    Linear & 256 & 1024 \\
    GELU & 1024 & 1024 \\
    Dropout & 1024 & 1024 \\
    Linear & 1024 & 256 \\
    GELU & 256 & 256 \\
    Dropout & 256 & 256 \\
    Linear & 256 & C \\
    \bottomrule
  \end{tabular}
\end{table}

\section{Evaluation metrics}
\paragraph{Accuracy}
Accuracy is defined as the ratio of correctly predicted labels to the total number of samples:
\begin{equation}
\text{Acc} = \frac{N_{\text{cor}}}{N_{\text{total}}},
\end{equation}
where $N_{\text{cor}}$ is the number of correctly classified faces or shapes, and $N_{\text{total}}$ is the total number of labels in the evaluation set (e.g., total number of faces for face-level tasks, or total number of models for classification).
\paragraph{Mean intersection over union}
To evaluate per-face prediction quality, especially under class imbalance, we use the mean intersection over union (mIoU). For a given class $c$, let $A_c$ be the set of predicted labels and $B_c$ the ground-truth labels. The IoU for class $c$ is defined as:
\begin{equation}
    \text{IoU}_c = \frac{|A_c \cap B_c|}{|A_c \cup B_c|}.
\end{equation}

The overall mIoU is the average over all $N_c$ classes:
\begin{equation}
\text{mIoU} = \frac{1}{N_c} \sum_{c=1}^{N_c} \text{IoU}_c.
\end{equation}
This metric measures the overlap between predicted and true labels and reflects both precision and recall at the class level.

\section {Pre-training Convergence Analysis}
\begin{figure}[h]
    \centering
    
    \includegraphics[width=0.8\textwidth]{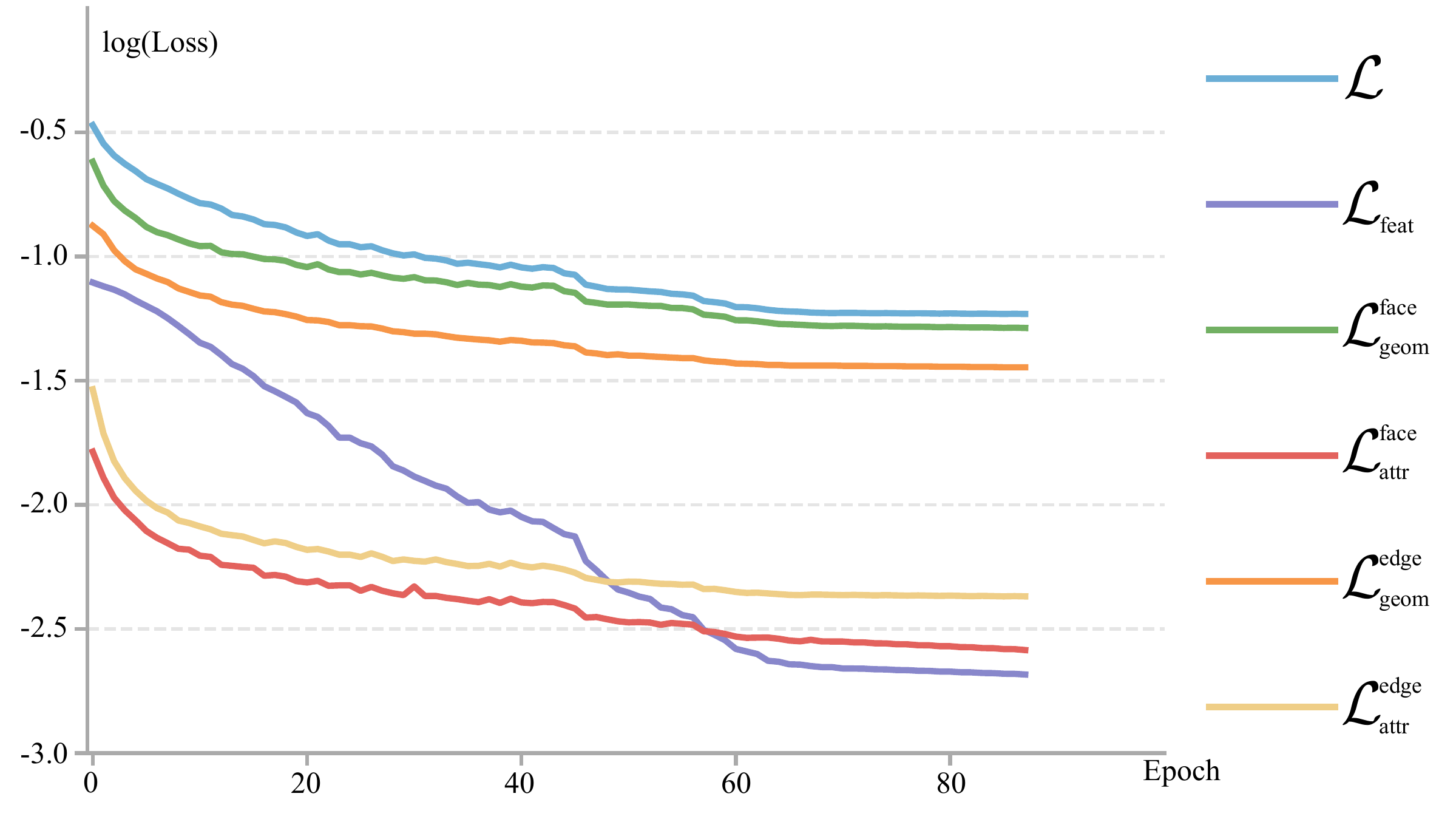} 
    \caption{The pre-training loss convergence curves. The y-axis denotes the logarithmic loss, showing the total loss and its five individual components over the training epochs.}
    \label{fig:loss_curve}
\end{figure}

To detail the pre-training optimization process, we visualize the convergence curves of the reconstruction losses. As shown in Fig.~\ref{fig:loss_curve}, the y-axis represents the logarithmic loss value ($\log(\text{Loss})$), and the x-axis represents the training epochs. The figure plots the total loss $\mathcal{L}$ alongside its five individual components: latent feature $\mathcal{L}_{feat}$, face geometry $\mathcal{L}_{geom}^{face}$, face attribute $\mathcal{L}_{attr}^{face}$, edge geometry $\mathcal{L}_{geom}^{edge}$, and edge attribute $\mathcal{L}_{attr}^{edge}$.

Throughout the training process, all loss components decrease steadily. In the later epochs, all curves gradually plateau. This consistent trend indicates that the parallel decoding branches converge stably without exhibiting divergence during the masked autoencoding task.

\section{Computational Cost Analysis}

To provide details on the computational cost of the proposed method, we report the model parameters, floating-point operations (FLOPs), inference latency, and throughput. The profiling is conducted using CAD models with an average of 30 faces on a single NVIDIA A800 GPU, consistent with the hardware setup described in the main paper. 

As shown in Table~\ref{tab:comp_cost}, we report the metrics separately for the self-supervised pre-training stage (comprising the full masked autoencoder architecture) and the downstream task stage (comprising the pre-trained encoder and the task-specific network).

\begin{table}[h]
\centering
\caption{Computational cost profiling for the pre-training and downstream stages. Profiling is based on CAD models with an average of 30 faces.}
\label{tab:comp_cost}
\resizebox{\textwidth}{!}{
\begin{tabular}{lcccc}
\toprule
\textbf{Stage} & \textbf{Parameters (M)} & \textbf{FLOPs (G)} & \textbf{Latency (ms)} & \textbf{Throughput (models/s)} \\
\midrule
Pre-training & 8.15 & 20.98 & 25.99 & 38.47 \\
Downstream   & 6.09 & 5.26  & 10.42 & 96.01 \\
\bottomrule
\end{tabular}
}
\end{table}

\textbf{Pre-training Cost:} During the pre-training stage, the full architecture contains 8.15M parameters and requires 20.98G FLOPs per forward pass. The average latency is 25.99 ms per model, corresponding to a throughput of 38.47 models per second, which constitutes a one-time computational cost.

\textbf{Downstream Cost:} For downstream tasks, the pre-training decoder is replaced by specific task networks tailored for the respective applications. The adapted downstream network comprises 6.09M parameters and requires 5.26G FLOPs. The latency in this stage is 10.42 ms per model, with a throughput of 96.01 models per second.

\textbf{Scalability Considerations:}It is worth noting that these profiling results are based on typical CAD models averaging 30 faces. When extending this framework to massive industrial assemblies with significantly more faces, the memory footprint and inference latency will naturally scale up.

\section{More Ablation Study}
\begin{table}[t]
\centering
\caption{More Ablation study on MFInstSeg under various supervision ratios.}
\label{tab:more_ablation}
\small
\setlength{\tabcolsep}{1.70pt}
\renewcommand{\arraystretch}{1.0}
\begin{tabular}{lccccccc}
\toprule
\textbf{Configuration} & \textbf{0.1\%} & \textbf{0.5\%} & \textbf{1\%} & \textbf{1.5\%} & \textbf{2\%} & \textbf{3\%} & \textbf{100\%} \\
\midrule
w/o Mask attr       & 84.48 & 94.15 & 95.42 & 96.34 & 96.60 & 97.20 & 99.41 \\
Feature all         & 76.77 & 90.86 & 93.79 & 94.67 & 95.72 & 96.56 & 99.33 \\
Geometry masked    & 78.03 & 91.14 & 93.91 & 95.23 & 95.81 & 96.71 & 99.37 \\
Mask ratio(50\%) &86.99 &96.02&97.18&97.43&97.60&98.02&99.49\\
Mask ratio(60\%) &87.41 &96.12&97.19&97.46&97.73&98.05&99.51\\
Mask ratio(80\%) &87.01 &95.66&96.86&97.37&97.62&97.96&99.46\\
\midrule
\textbf{Ours (Default)}          & \textbf{88.75} & \textbf{96.19} & \textbf{97.31} & \textbf{97.67} & \textbf{97.92} & \textbf{98.26} & \textbf{99.53} \\
\bottomrule
\end{tabular}

\end{table}

Due to the page limit of the main text, we provide additional ablation studies in this section to further validate our pre-training configurations and loss designs. All experiments in this section are conducted on the MFInstSeg dataset, following the identical training protocol described in the main paper. The quantitative results are summarized in Table~\ref{tab:more_ablation}.

\textbf{1. Masking Attributes vs. Geometry Only.} 
In our default setting, we mask both the raw geometries and the discrete attributes. To validate this, we test a configuration where only geometric information is masked, while attribute information remains fully visible to the network (\textit{w/o Mask attr}). As shown in Table~\ref{tab:more_ablation}, this results in a noticeable performance drop (e.g., from 88.75\% to 84.48\% at the 0.1\% supervision level). This demonstrates that keeping attributes visible provides the network with a trivial shortcut to infer local semantics, thereby weakening the information bottleneck and hindering the learning of deep, transferable representations.

\textbf{2. Supervision Domains for Loss Functions.}
We ablate the regions over which our pre-training losses are computed to justify our asymmetric design choices between feature and geometry supervisions:
\begin{itemize}
    \item \textit{Feature all:} By default, the latent feature loss ($\mathcal{L}_{feat}$) is computed strictly on the masked entities. If we force the network to compute this loss on all entities (both masked and unmasked), the accuracy drops drastically to 76.77\% at 0.1\% supervision. This is because unmasked features are already exposed to the encoder; supervising them encourages a trivial identity mapping, which distracts the network from its primary task of deducing missing structures.
    \item \textit{Geometry masked:} Conversely, our default explicit geometry losses ($\mathcal{L}_{geom}^{face}$, $\mathcal{L}_{geom}^{edge}$) are computed over the entire BRep model. If we restrict the geometry loss to only the masked regions (\textit{Geometry masked}), performance drops to 78.03\% at 0.1\% supervision. Reconstructing the full geometry acts as a strong global regularizer, ensuring that the decoder maintains overall shape consistency and spatial coherence rather than optimizing masked patches in isolation.
\end{itemize}

\textbf{3. Sensitivity to Masking Ratios.} 
We also evaluate the sensitivity of our framework to different pre-training masking ratios (\textit{Mask ratio 50\%}, \textit{Mask ratio 60\%}, and \textit{Mask ratio 80\%}). The results indicate that while our architecture remains relatively robust across different masking regimes, our default configuration~(with a 70\% masking ratio) yields the most optimal and balanced representation, consistently achieving the highest accuracy across all low-data settings.

\section{Additional Visualization Results}
\begin{figure*}[!t]
  \centering
  \includegraphics[width=0.95\textwidth]{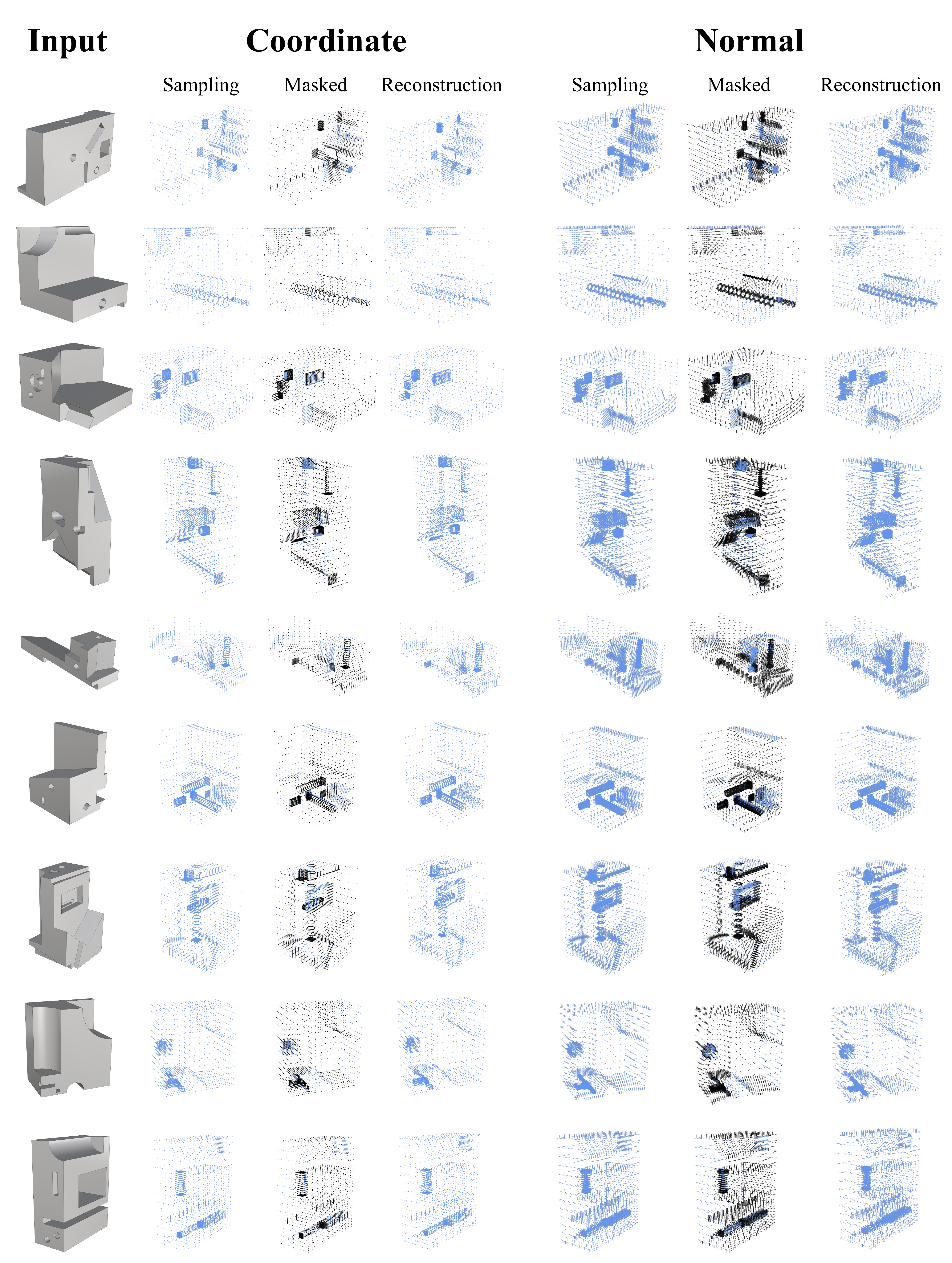}
  \vspace{-2mm}
  \caption{More reconstruction examples with face coordinates and face normals.}
  \label{fig:recon_more}
\end{figure*}

\begin{figure*}[!t]
  \centering
  \includegraphics[width=0.89\textwidth]{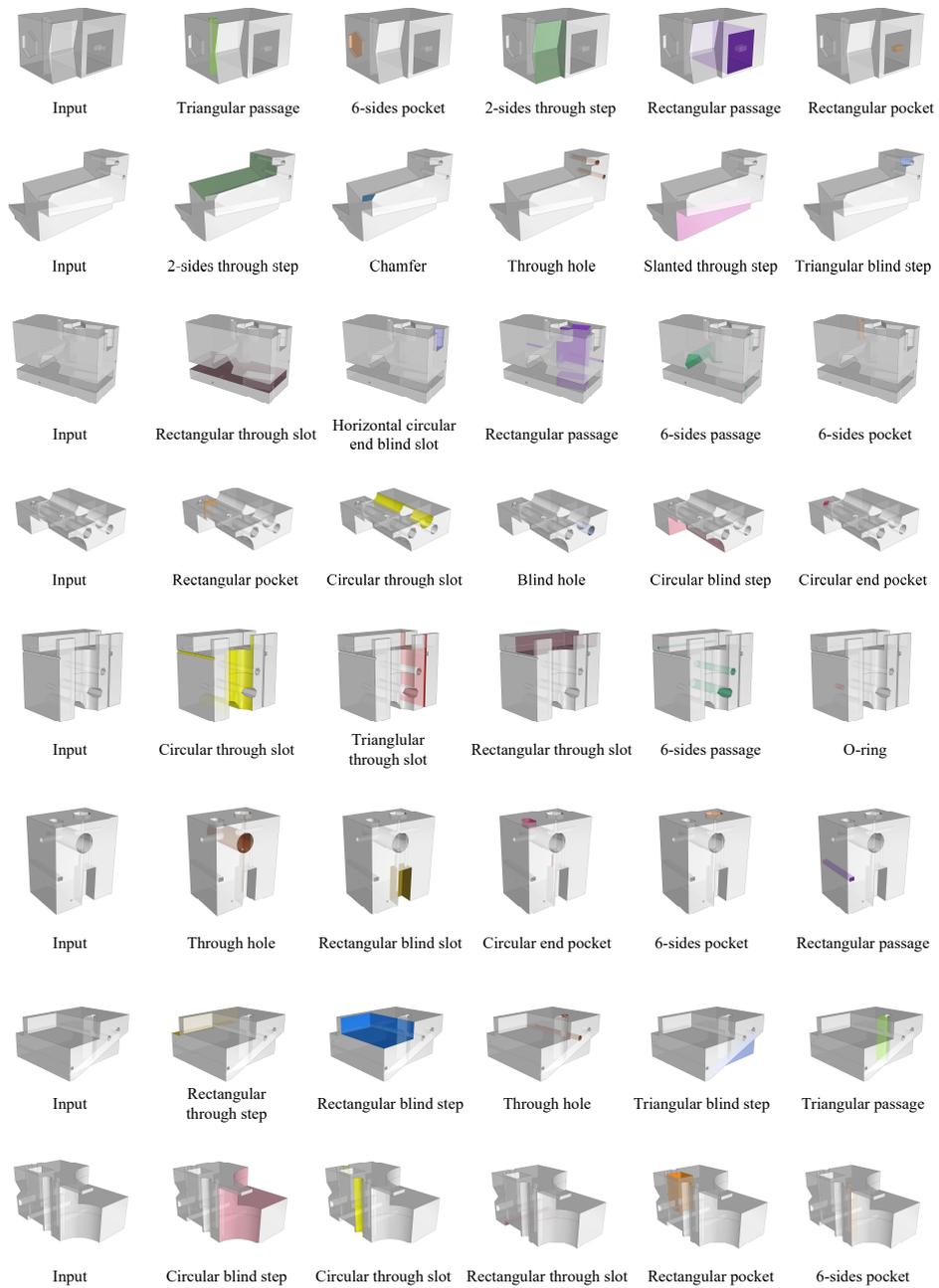}
  \vspace{-2mm}
  \caption{More results of machining feature recognition.}
  \label{fig:mfr_more}
\end{figure*}

\begin{figure*}[!t]
  \centering
  \includegraphics[width=0.89\textwidth]{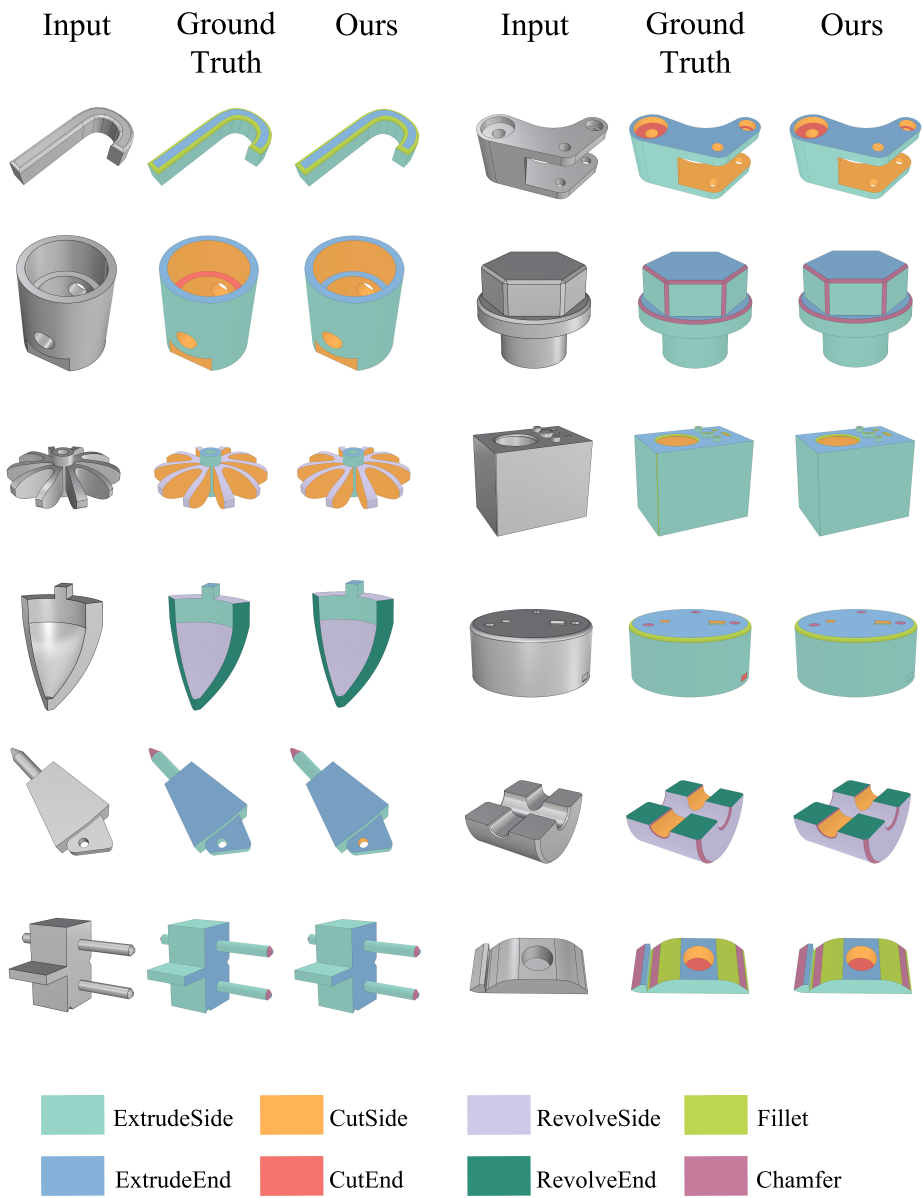}
  \vspace{-2mm}
  \caption{More results of modeling segmentation.}
  \label{fig:seg_more}
\end{figure*}

To provide a more comprehensive qualitative evaluation, we present additional visualization results in this section. Fig.~\ref{fig:recon_more} displays more reconstruction examples, specifically visualizing the face coordinates and face normals, which further demonstrates the effectiveness of our masked pre-training strategy. Regarding downstream tasks, we show extended results for machining feature recognition in Fig.~\ref{fig:mfr_more} and modeling segmentation in Fig.~\ref{fig:seg_more}. These visualizations confirm that our pre-trained encoder learns robust and generalizable representations for diverse CAD tasks.

\end{document}